\documentclass{article}

\usepackage[a4paper]{geometry}
\usepackage{graphicx}
\usepackage[textwidth=8em,textsize=small]{todonotes}
\usepackage{amsmath}
\usepackage{natbib}
\usepackage{verbatim}
\usepackage{booktabs}
\usepackage{url}
\usepackage{amssymb}
\usepackage{graphicx}
\usepackage{setspace}
\usepackage{wasysym}
\usepackage{enumerate}
\usepackage{enumitem}
\usepackage{float}
\usepackage[plain,ruled]{algorithm2e}
\usepackage{comment}
\usepackage{tikz}
\usepackage{graphicx}
\usepackage{array}
\usepackage{authblk}
\usetikzlibrary{shapes,decorations,arrows,calc,arrows.meta,fit,positioning}
\tikzset{
    -Latex,auto,node distance =1 cm and 1 cm,semithick,
    state/.style ={ellipse, draw, minimum width = 0.7 cm, fill = white!25},
    point/.style = {circle, draw, inner sep=0.04cm,fill,node contents={}},
    bidirected/.style={Latex-Latex,dashed},
    el/.style = {inner sep=2pt, align=left, sloped}
}

\newcommand{\bs}{\boldsymbol{s}}
\newcommand{\E}{\text{E}}

\newcommand{\T}{\mathrm{\scriptscriptstyle T}}

\newcommand{\ipw}{\mathrm{ipw}}

\newcommand{\bc}{\mathrm{bc}}

\newcommand{\lp}{\mathrm{lp}}
\newcommand{\loss}{\mathrm{loss}}

\newcommand{\mP}{\mathbb{P}}
\usepackage{comment}
\newtheorem{theorem}{Theorem}\newtheorem{definition}{Definition}\newtheorem{remark}{Remark}\newtheorem{example}{Example}
\newcommand{\indep}{\perp \!\!\! \perp}
\newcommand*\samethanks[1][\value{footnote}]{\footnotemark[#1]}

\title{Estimating spatially varying health effects of wildland fire smoke using mobile health data}
\author[1]{Lili Wu\thanks{Joint first authors}}
\author[1]{Chenyin Gao\samethanks[1]}
\author[1]{Shu Yang\thanks{Corresponding Author: Department of Statistics, North Carolina State University, 2311 Stinson Dr. Raleigh, NC 27695, Raleigh, U.S.A. \\ \sffamily E-mail: syang24@ncsu.edu}}
\author[1]{Brian J. Reich}
\author[2]{Ana G. Rappold}
\affil[1]{Department of Statistics, North Carolina State University, Raleigh, U.S.A.}
\affil[2]{Environmental Protection Agency, Research Triangle Park, Durham, U.S.A.}

 \usepackage{xr}
\makeatletter
\newcommand*{\addFileDependency}[1]{
  \typeout{(#1)}
  \@addtofilelist{#1}
  \IfFileExists{#1}{}{\typeout{No file #1.}}
}
\makeatother	
\newcommand*{\myexternaldocument}[2]{%
    \externaldocument[#1]{#2}%
    \addFileDependency{#2.tex}%
    \addFileDependency{#2.aux}%
}
\myexternaldocument{supple-}{supple}

\begin{document}
	\maketitle
	\date{}
\begin{abstract}
Wildland fire smoke exposures are an increasing threat to public health, and thus there is a growing need for studying the effects of protective behaviors on reducing health outcomes. Emerging smartphone applications provide unprecedented opportunities to deliver health risk communication messages to a large number of individuals when and where they experience the exposure and subsequently study the effectiveness, but also pose novel methodological challenges. Smoke Sense, a citizen science project, provides an interactive smartphone app platform for participants to engage with information about air quality and ways to protect their health and record their own health symptoms and actions taken to reduce smoke exposure. We propose a new, doubly robust estimator of the structural nested mean model parameter that accounts for spatially- and time-varying effects via a local estimating equation approach with geographical kernel weighting. Moreover, our analytical framework
is flexible enough to handle informative missingness by inverse probability
weighting of estimating functions. We evaluate the new method using extensive simulation studies and apply it to Smoke Sense data reported by the citizen scientists to increase the knowledge base about the relationship between health preventive measures and improved health outcomes. Our results estimate how the protective behaviors' effects vary over space and time and find that protective behaviors have more significant effects on reducing health symptoms in the Southwest than the Northwest region of the USA.

\end{abstract}
 \vfill{}

\section{Introduction\label{sec:Introduction}}
Wildland fire smoke is an emerging health issue as one of the largest
sources of unhealthy air quality, attributing an estimated $340,000$
excess deaths each year globally \citep{johnston2012estimated}. Although
there are a number of exposure-reducing actions recommended, there
is a lack of evidence of the long-term reduction in the number of
adverse health outcomes by taking health-protective behaviors (treatments). The Smoke Sense citizen
science initiative \citep{rappold2019smoke}, introduced by the researchers
at the Environment Protection Agency, aims to engage citizen scientists
and develop a personal connection between changes in environmental
conditions and changes in personal health to promote health-protective
behavior during wildland fire smoke {exposure}. The overarching objective of
the Smoke Sense project is to develop and maintain an interactive
platform for building knowledge about wildfire smoke, health, and
protective actions to improve public health outcomes. The use of smartphone
application (app) is designed as a risk reduction intervention based
on the theory of planned behavior and health belief model. Through
Smoke Sense, participants can report their perceptions of risk, adoptions
of protective health behaviors and health symptoms. Therefore, Smoke Sense is uniquely
placed to address this knowledge gap.

App-based platforms provide unprecedented opportunities to reach users
and learn about the personal motivations to engage with the information
delivered through the apps. However, the data also present analytical
and methodological challenges as summarized below. (i) Adoption of
health-protective behaviors (treatments) were left to the participants
and may depend on the participant's characteristics and perceptions
of benefits and barriers of these actions. Self-selection can potentially
result in confounding by indication \citep{pearl2009causality}. Thus,
statistical methods must adequately adjust for participant characteristics
that confound the relationship between their behaviors and the outcome.
This challenge is more pronounced in longitudinal Smoke Sense data
because of the time-varying treatments and confounding. (ii) Although
participants can be viewed as independent samples, the causal effect
of treatment may vary over the study's large and socially- and environmentally-diverse
domain that is not explained by the observed covariates, and existing causal methods typically assume the structural treatment effect model accounting for the observed covariates is homogeneous. (iii) Participants were more likely to self-report
when they experienced smoke or had health symptoms, leading to informative
non-responses \citep{rubin1976inference}; i.e., the missingness mechanism
due to non-responses depends on the missing values themselves even
adjusting for all observed variables. Failure to appropriately account
for informative missingness may also lead to bias. These opportunities
and challenges are identified for the Smoke Sense citizen science
platform; however, a causal inference framework to study the relationship
between interventions and health outcomes from mobile application
data would have wider application in health and behavior research.

Confounding by indication poses a unique challenge to drawing valid
causal inference of treatment effects from observational studies.
For example, sicker patients are more likely to take the active treatment,
whereas healthier patients are more likely to take the control treatment.
Consequently, it is not meaningful to compare the outcome from the
treatment group and the control group directly. Moreover, in longitudinal
observational studies, confounding by indication is likely to be time-dependent
\citep{robins2009estimation}, in the sense that time-varying prognostic
factors of the outcome affect the treatment assignment at each time,
and thereby distort the association between treatment and outcome
over time. In these cases, the traditional regression methods are
biased even after adjusting for the time-varying confounders \citep{robins1992g}.

Parametric g-computation \citep{robins1986new}, Marginal Structural
Models (\citealt{robins2000marginal}), and Structural Nested Models
(\citealp{robins1992g}) are three major approaches to overcoming
the challenges with time-varying confounding in longitudinal observational
studies. However, the existing causal models typically assume the spatial homogeneity
of the structural treatment effect models accounting for the observed covariates; i.e., the treatment effect is a constant
across locations. This assumption is questionable in studies with
smartphone applications, including the Smoke Sense Initiative, where
the smoke exposure, study participant's motivations, and treatment
vary across a large, socially, and environmentally diverse domain.
It is likely that the treatment effect varies across spatial locations.
Although spatially varying coefficient models exist \citep[e.g., ][]{gelfand2003spatial},
they restrict to study the associational relationship of treatment
and outcome and thus lack causal interpretations. \cite{reich2021review} provided a comprehensive review of spatial causal inference methods and suggested that causal models with spatially varying effects are largely needed.  

We establish the causal effect model that allows the causal
effect to vary over space accounting for unmeasured spatial treatment effect modifiers. Under the standard sequential
randomization assumption, we show that the local causal parameter
can be identified based on a class of estimating equations. To borrow
information from nearby locations, we adopt the local estimating equation
approach via local polynomials (\citealp{fan1996local}) and geographical
kernel weighting (\citealp{fotheringham2003geographically}). Moreover,
we also derive the asymptotic theory and propose an easy-to-implement
inference procedure based on the wild bootstrap. Within the new framework,
a challenge arises for selecting the bandwidth parameter determining
the scale of spatial treatment effect heterogeneity. Existing methods
rely on cross-validation on predictions, where a typical loss function
is the mean squared prediction error, which is not applicable under
the causal framework because the task is estimating causal effects
rather than predicting outcomes. This is due to the fundamental problem
in causal inference that not all ground-truth potential outcomes can
be observed \citep{holland1986statistics}. We propose a loss function
using a new balancing criterion for bandwidth selection. Finally,
we propose to use an instrumental variable for the Smoke Sense application
that adjusts for informative missingness.

Our analytic framework is appealing for multiple reasons. First, the
framework is semiparametric and does not require modeling the full
data distribution. Second, it is doubly robust in the sense that,
with a correct treatment effect model, the proposed estimator is consistent
if either the propensity score model or a nuisance outcome mean model
is correctly specified. Third, it is flexible enough to handle informative
missingness by inverse probability weighting of estimating functions.
Fourth, it is a very general framework of spatially- and time-varying
causal effect estimation which has much potential in many other mobile
health applications such as diagnostic and treatment support, disease
and epidemic outbreak tracking, etc \citep{adibi2014mhealth}.

The rest of the paper is organized as follows. Section \ref{sec:Basic-setup}
introduces the data sources and notation. Section \ref{sec:Global Structural Nested Mean Models}
describes existing global structural nested mean models (SNMMs). Section
\ref{sec:Local Structural Nested Means Model} develops new local
SNMMs, local estimation, and the asymptotic properties. We extend
the framework to handle informative non-responses with instrument
variables in Section \ref{sec:non-responses}. We apply the method
to the simulated data and real data collected from the Smoke Sense
Initiative in Section \ref{sec:simulation} and Section \ref{sec:Real data},
respectively. We conclude the article with a discussion in Section
\ref{sec:Discussion}.

\section{Smoke Sense citizen science study \label{sec:Basic-setup}}

The dataset from the Smoke Sense citizen science study combines the
self-reported observations of smoke, health symptoms, and behavioral
actions taken in response to smoke and the estimated exposure to wildfire
smoke recorded by the National Oceanic and Atmospheric Administration's
Office of Satellite and Product Operations Hazard Mapping System's
Smoke Product (HMS).

\subsection{Smoke Sense app}

\begin{figure}[ht!]
\centering{ \includegraphics[width=0.5\textwidth]{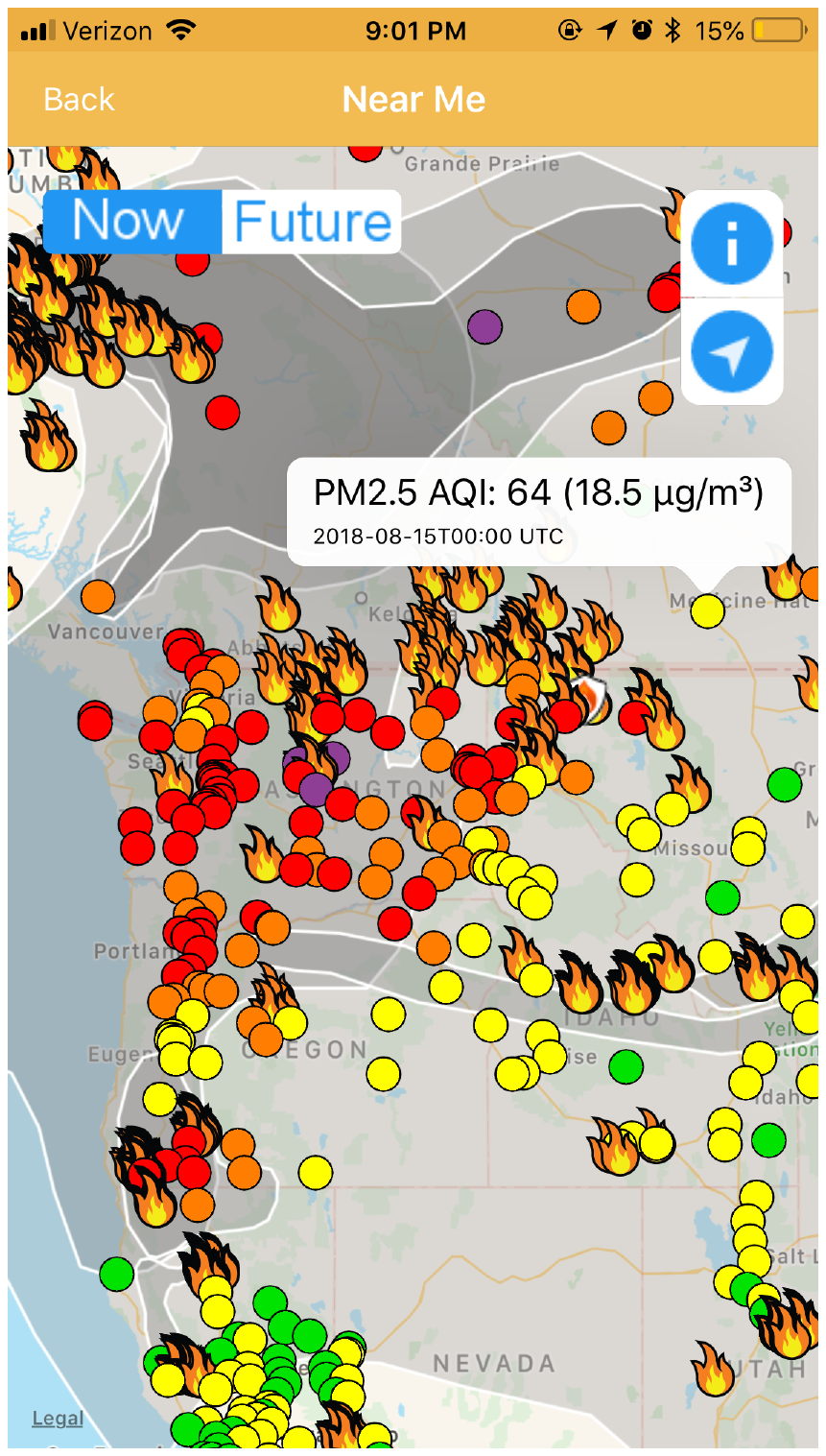}} \caption{\label{fig:Data} A screen shot of the Smoke Sense App alerting the
user of local fires and air quality.}
\end{figure}

The Smoke Sense citizen science study is facilitated through the use
of a smartphone application, a publicly available mobile application
on the Google Play Store and the Apple App Store. The app invites
users to record their smoke observations and health symptoms, as well
as the actions they took to protect their health. In the app, participants
can also explore current and forecasted daily air quality, learn where
the current wildfires are burning (Figure \ref{fig:Data}), read about
the progress of the wildfire suppression efforts, and observe satellite
images of smoke plumes. Participants are also invited to play educational
trivia games, explore what other users are reporting, learn strategies
to minimize exposure, and learn about the health impacts of wildland
fire smoke. Participants earn badges for the level of participation
as users, observers, learners, and reporters. 

In this study, the outcome of interest is the number of user-reported
adverse health symptoms during the 2019 smoke season, reported as
weekly summaries. The weekly number of adverse events ranged from $0$ to $15$ with a mean (sd) of $2.6$ ($3.2$). The treatment is
a binary indicator of whether the participant took strong protective
behaviors -- staying indoors with extra protective behaviors like
using an air cleaner or a respirator mask. Other variables include
the baseline information when registered in the app including age,
gender, first 3-digits of the zip code, etc, and time-varying variables
including self-reported smoke experience, days of visibility impacted,
etc, which users are reminded of reporting every week. More details
about the variables are provided in Supplementary Materials. In our
analysis, we include $n=1882$ users who reported baseline and time-varying
variables. Among these users, $471$ reported more than once and the
maximum number of reporting is $61$.

\subsection{Wildfire smoke exposure\label{subsec:smoke_exposure}}

\begin{figure}[!ht]
\centering{ \includegraphics[width=\textwidth]{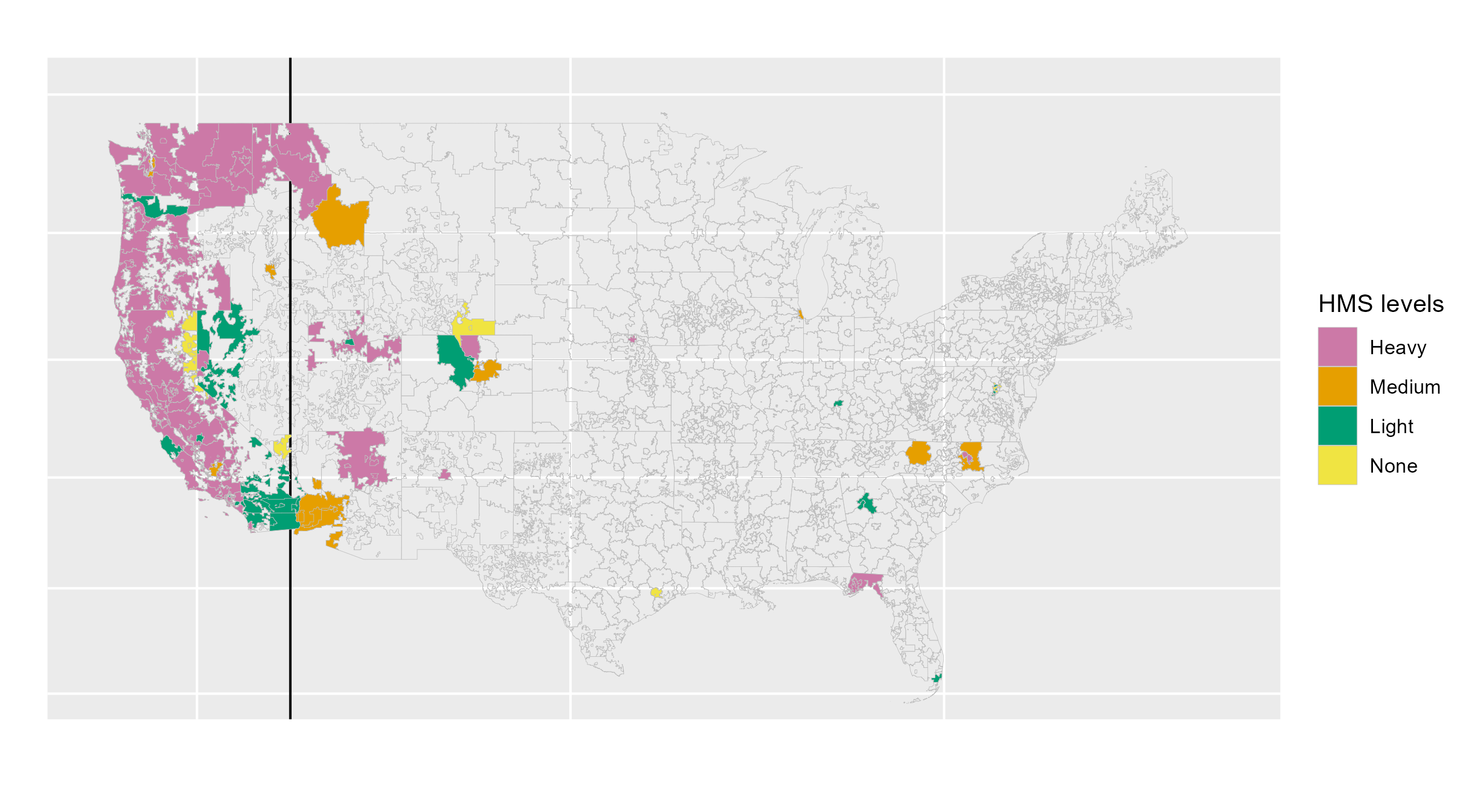}}
 \caption{Maximum daily HMS smoke density level in each three-digit zip code
over the study period. The smoke density has four levels: None, Light,
Medium, and Heavy (corresponding to 0, 5, 16, and 27 $\mu g/m^{3}$,
respectively). The vertical line is at longitude -115.}
\label{fig:map_hms} 
\end{figure}

For each user-week, the exposure to smoke is determined based on HMS
(\url{http://satepsanone.nesdis.noaa.gov/pub/volcano/FIRE/HMS\_ARCHIVE/}).
The HMS data contains the spatial contours of satellite observed imagery
of smoke together with the estimated density of fine particulate matter
(${\rm PM}_{2.5}$) based on the air quality model. The smoke density in the HMS data is summarized by four levels: none, light (${\rm PM}_{2.5}$ range:
$0-10$ $\mu g/m^{3}$, medium ($10.5-21.5$ $\mu g/m^{3}$), and
dense ($\geq22\mu g/m^{3}$), where each level is summarized by the
midpoint of the corresponding range (0, 5, 16, and 27 $\mu g/m^{3}$,
respectively). We map the highest daily exposure value to each zip
code and aggregate the HMS data by taking the maximum value over the
3-digit zip code (zip3). We restrict our analysis to the time period
from September 2018 to the end of 2019 and across zip3 where there
were more than 10 smoke sense users. Figure \ref{fig:map_hms} shows
the map of the maximum weekly smoke density across zip3 geographic
locations during the study period, where the west coast showed heavier
wildland fire smoke occurrence.

\begin{figure}
\centering{ \includegraphics[width=1\textwidth]{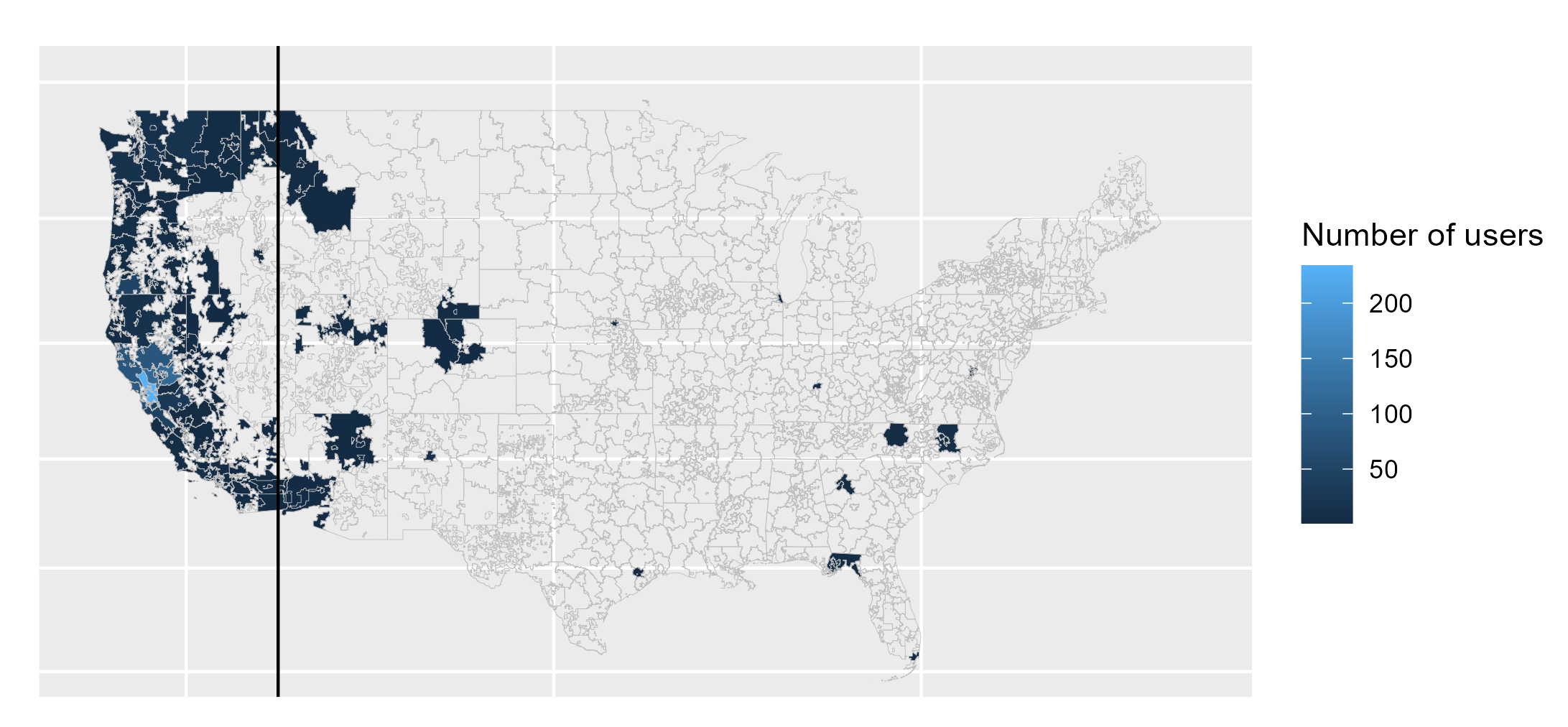}}
 \caption{\label{fig:map_num_users}The number of users in each three-digit
zip code over the study period. The vertical line is at longitude
-115.}
\end{figure}

\subsection{Notation\label{subsec:notation}}

We follow the notation from the standard structural nested model literature
\citep{robins1994correcting}. We assume $n$ subjects are monitored
over time points $t_{0},\ldots,t_{K}$. For privacy concerns, the
spatial location of the subjects are summarized by their 3-digit zip
codes. To obtain the latitude and longitude of the spatial location,
$\bs=(s_{1},s_{2})$, we average the latitude and longitude coordinates
over all the 5-digit zip codes which have the same 3-digit zip codes
(the latitude and longitude coordinates corresponding to each 5-digit
zip code are available at \url{https://public.opendatasoft.com/explore/dataset/us-zip-code-latitude-and-longitude/export/}).
The subjects are assumed to be an independent sample, which is a plausible assumption since the health symptoms in the data are not like infectious decease so that the interference is not likely. And for simplicity, we omit a subscript for subject and location; that is, any variable will have a subject index $i$ and a location index $\boldsymbol{s}$ implicitly. Let $A_{k}$ be the binary treatment
at $t_{k}$ ($A_{k}=1$ if the subject reported taking the strong
protective behaviors, and $A_{k}=0$ if the subject did not take any
behavior or took some mild protective behaviors). Let $X_{0}$ be
a vector of baseline variables, including the subject's demographic
information (e.g., sex, age, race, education level), baseline health
information (e.g., pre-existing conditions, physical activity level,
time spent outdoors), and current beliefs about smoke and air pollution.
Let $X_{k}\in\text{\ensuremath{\mathcal{X}_{k}}}$ be a vector of
baseline covariates and time-varying variables at $t_{k}$ (e.g.,
the recent experience of smoke, feeling status, visibility). Let $Y_{k}$
be the outcome at $t_{k}$ (the total number of symptoms the subject
had including e.g. anxiety, asthma attack, chest pain).

We use overbars to denote a variable's history; e.g., $\overline{A}_{k}=\{A_{m}:m=0,\ldots,k\}$,
and the complete history $\overline{A}_{K}$ abbreviates to $\overline{A}$.
Let $Y_{k}^{(\overline{a})}$ be the outcome at $t_{k}$, possibly counterfactual,
had the subject followed treatment regime $\overline{a}$ over the study
period from $t_{0}$ to $t_{K}$. For simplicity, we use $Y_{k}^{(\overline{a}_{m})}$
$(m\le k)$ to denote the potential outcome at $t_{k}$ had the subject
followed treatment regime $\overline{a}_{m}$ until $t_{m}$ and no treatment
onwards. We assume that the observed outcome $Y_{k}$ is equal to
$Y_{k}^{(\overline{A})}$ for $k=0,\ldots,K$ as different strong protective behaviors considered in our study presumably reduce exposure to smoke or harmful air to a similar extent. Finally, $V=(\overline{A},\overline{X},\overline{Y})$
denotes the subject's full records. Up to Section \ref{sec:non-responses},
we shall assume that all subjects' full records are observed. We let
$\mP$ be the probability measure induced by $V$ and $\mP_{n}$ be
the empirical measure for $V_{1},...,V_{n}$; i.e., $\mP_{n}f(V)=n^{-1}\sum_{i=1}^{n}f(V_{i})$
for any real-valued function $f(v)$.

\section{Global structural nested mean models\label{sec:Global Structural Nested Mean Models}}

The treatment effect is defined in terms of the expected value of potential outcomes under different treatment trajectories. Because both the treatment and outcome are longitudinal, and there may be a lag time between the treatment and its effect, we define the causal effect of the treatment at $t_{m}$ on the outcome at $t_{k}$ for $k\ge m$. SNMMs are one type of causal models designed to properly handle such time-varying treatments and confounders. We have provided additional details using a \textit{directed acyclic graph} (DAG) in the Supplementary Material to illustrate their ability to capture time-varying treatment effects, as compared to standard regression models.

\begin{definition}[Global SNMM]\label{assumpM-0}
Let $\overline{a}_{-1}$ denote a null set by convention, and $\gamma_{m,k}(\psi^{*})=\gamma_{m,k}(\overline{a}_{m},\overline{x}_{m};\psi^{*})$
be a known function of $(\overline{a}_{m},\overline{x}_{m})$ with a vector of unknown parameters $\psi^{*}\in\mathcal{R}^{p}$ with a fixed $p\geq1$. For $0\leq m\leq k\leq K$,
the treatment effect is characterized by 
\begin{equation}
\E\left\{ Y_{k}^{(\overline{a}_{m-1},a_{m})}-Y_{k}^{(\overline{a}_{m-1},0)}\mid\overline{A}_{m-1}=\overline{a}_{m-1},\overline{X}_{m}=\overline{x}_{m}\right\} =\gamma_{m,k}(\psi^{*}),\label{eq:trt model}
\end{equation}
where the causal effect is the expected difference in the response
at $t_{k}$ between two counterfactual regimes with the same treatment
before $t_{m}$, different treatment at $t_{m}$, and no treatments
after $t_{m}$.

\end{definition}

To help understand the model, consider the following example.

\begin{example}\label{eg:linear-SNMM-1} 
Assume $\gamma_{m,k}(\psi^*)=\delta
\exp\left\{
-(t_k-t_m-\mu)^2/(2\sigma^2)
\right\}a_m$, where $\psi^*=(\delta,\mu,\sigma^2)$ is the vector of parameters.
\end{example}
This model entails a key feature for the treatment
effect curve: the
effect of treatment $a_{m}$ on future health outcomes increases smoothly
over time, reaches to a peak effect $\delta$ after $\mu$ units of time, and decreases
to zero as time further increases. The rationale behind this delayed effect pattern is that taking protective behaviors may not lead to immediate changes in health outcomes. Therefore, the impact on the total number of adverse health symptoms is anticipated to peak after a certain duration and gradually decrease to zero for large time lags.

This framework can also include effect modifiers. For example, we
can consider $\gamma_{m,k}(\psi^{*})=a_{m}(1,x_{m}^{\T})\psi^{*}$,
where $x_{m}$ is a $(p-1)$-vector of the individual characteristics
and $\psi^{*}=(\psi_{1}^{*},\ldots,\psi_{p}^{*})^{\T}$. Therefore,
the class of SNMMs has important applications in precision medicine
\citep{chakraborty2013statistical} for the discovery of optimal treatment
regimes that are tailored to individuals' characteristics and environments.

Parameter identification requires the typical sequential randomization
assumption \citep{robins1992g} that for $0\leq m\leq k\leq K$, $Y_{k}^{(\overline{a}_{m})}\indep A_{m}\mid\overline{L}_{m}$,
where $\overline{L}_{m}$$=(\overline{A}_{m-1},\overline{X}_{m},\overline{Y}_{m-1})$.
This assumption holds if $\overline{L}_{m}$ captures all confounders for
the treatment at $t_{m}$ and ensuing outcomes. Define the propensity
score as $e(\overline{L}_{m})=\mathrm{P}(A_{m}=1\mid\overline{L}_{m})$. Moreover,
define $H_{k}^{(\overline{A}_{m-1})}(\psi^{*})=Y_{k}-\sum_{l=m}^{k}\gamma_{l,k}(\psi^{*})$
and $\mu_{m,k}(\overline{L}_{m})=\E\{H_{k}^{(\overline{A}_{m-1})}(\psi)\mid\overline{L}_{m}\}$.
Intuitively, $H_{k}^{(\overline{A}_{m-1})}(\psi^{*})$ removes the accumulated
treatment effects from $t_{m}$ to $t_{k}$ from the observed outcome
$Y_{k}$, so it mimics the potential outcome $Y_{k}^{(\overline{a}_{m-1})}$
had the subject followed $\overline{a}_{m-1}=\overline{A}_{m-1}$ but no treatment
onwards. The sequential randomization assumption states that $Y_{k}^{(\overline{a}_{m-1})}$
and $A_{m}$ are independent given $\overline{L}_{m}$. \citet{robins1992g}
showed that $H_{k}^{(\overline{A}_{m-1})}(\psi^{*})$ inherits this property
in the sense that $\E\{H_{k}^{(\overline{A}_{m-1})}(\psi)\mid A_{m},\overline{L}_{m}\}=\E\{H_{k}^{(\overline{A}_{m-1})}(\psi)\mid\overline{L}_{m}\}$.
As a result, with any measurable, bounded function $q_{k,m}:\overline{\mathcal{\mathcal{L}}}_{m}\rightarrow\mathcal{R}^{p}$,
\begin{equation}
G(V;\psi)=\sum_{m=1}^{K}\sum_{k=m}^{K}q_{k,m}(\overline{L}_{m})\left\{ H_{k}^{(\overline{A}_{m-1})}(\psi)-\mu_{m,k}(\overline{L}_{m})\right\} \{A_{m}-e(\overline{L}_{m})\}\label{eq:opt ef}
\end{equation}
is unbiased at $\psi^{*}$. Then, under a regularity condition that
$\E\{\partial G(V;\psi)/\partial\psi\}$ is invertible, the solution
to $\E\{G(V;\psi)\}=0$ uniquely exists, and therefore $\psi^{*}$
is identifiable.

The estimating function $G(V;\psi)$ depends on $q_{k,m}(\overline{L}_{m})$.
The choice of $q_{k,m}(\overline{L}_{m})$ does not affect the unbiasedness
but estimation efficiency. We adopt an optimal form of $q_{k,m}(\overline{L}_{m})$
given in the Supplementary Material. With this choice, the solution
to $\mathbb{P}_{n}G(V;\psi)=0$ has the smallest asymptotic variance
compared to other choices \citep{robins1994correcting}.

\section{Local structural nested mean model\label{sec:Local Structural Nested Means Model}}
Global SNMMs in Section \ref{sec:Global Structural Nested Mean Models}
allow time-varying treatment effects but not spatially varying treatment
effects. That is, the treatment effects for a given time are the same across space. Although the global SNMMs can model the spatial heterogeneity of treatment effects by using the observed spatial covariates, there may be 
unobserved location-specific heterogeneity. To overcome this issue, 
we extend to a new class of models that allows modeling spatial
treatment effect heterogeneity using spatially local parameters. Specifically, Section \ref{sec:spatial_SNMM} proposes the spatially varying SNMM to estimate the location-specific treatment effect. Section \ref{subsec:geo-weighted} introduces an extension of the location-specific estimating equations by incorporating the neighborhood information through geographically weighted regression. Section \ref{subsec:asymptotic} examines the asymptotic properties of the proposed estimator, and Sections \ref{subsec:bandwidth} to \ref{subsec:wild} offer guidance on tuning the bandwidth, bias correction, and inference in practice.

\subsection{Spatially varying structural nested mean models \label{sec:spatial_SNMM}}

The local SNMM is formulated by considering a vector of spatially varying parameters, denoted as $\psi^*(\boldsymbol{s})$, for each location $\boldsymbol{s}$ in (\ref{eq:trt model}).
\begin{definition}[Local SNMM]\label{assumpM} Let $\gamma_{m,k}\{\psi^{*}(\bs)\}=\gamma_{m,k}\{\overline{a}_{m},\overline{x}_{m};\psi^{*}(\bs)\}$
be a known function of $(\overline{a}_{m},\overline{x}_{m})$ with a vector
of spatially varying parameters $\psi^{*}(\bs)\in\mathcal{R}^{p}$
with fixed $p\geq1$. For $0\leq m\leq k\leq K$,
the treatment effect is characterized by 
\begin{equation}
\E\left\{ Y_{k}^{(\overline{a}_{m-1},a_{m})}-Y_{k}^{(\overline{a}_{m-1},0)}\mid\overline{A}_{m-1}=\overline{a}_{m-1},\overline{X}_{m}=\overline{x}_{m}\right\} =\gamma_{m,k}\{\psi^{*}(\bs)\},\label{eq:trt model-1}
\end{equation}
where $\boldsymbol{s}$ is a fixed location that is implicitly involved in the left-hand side of (\ref{eq:trt model-1}). 
\end{definition}

Consider the following example in parallel to Example \ref{eg:linear-SNMM-1}.

\begin{example}\label{eg:linear-SNMM}
Assume $\gamma_{m,k}\{\psi^*(\bs)\}=\delta(\bs) 
\exp[
-(t_k-t_m-\mu(\boldsymbol{s}))^2/\{2\sigma(\bs)^2\}]
a_m$, where $\boldsymbol{s}$ is a given spatial location.
\end{example} 
The local model in Example \ref{eg:linear-SNMM} entails that the treatment would increase the mean of the outcome for subjects at location $\bs$ by $\delta(\bs)$ at its peak when $t_{k} = t_{m}+\mu(\bs)$ if the subject had received the treatment at $t_{m}$. The spatial variations in the timing and intensity of peak effects are captured by $\mu(\bs)$ and $\delta(\bs)$, which can be attributed to various location-specific heterogeneitity.


\subsection{Geographically weighted local polynomial estimation \label{subsec:geo-weighted}}

There are an infinite number of parameters because $\psi^{*}(\bs)$
varies over $\bs$. Estimation of $\psi^{*}(\bs)$ at a given $\bs$
may become unstable with only a few observations at $\bs$, or even
infeasible at locations without any observations. To make estimation
feasible, one can make some global structural assumptions about $\psi^{*}(\bs)$
with a fixed number of unknown parameters. However, this approach
is sensitive to model misspecification. To overcome this difficulty,
we combine the ideas of local polynomial approximation and geographically
weighted regression. That is, we leave the global structure of $\psi^{*}(\bs)$
unspecified but approximate $\psi^{*}(\bs)$ locally by polynomials
of $\bs$. Then, we use geographical weighting to estimate the local
parameters by pooling nearby observations whose contributions diminish
with geographical distance.

To be specific, we consider estimating $\psi^{*}(\bs^{*})$ at a given
$\bs^{*}.$ We approximate $\psi^{*}(\bs)=\{\psi_{1}^{*}(\bs),\ldots,\psi_{p}^{*}(\bs)\}^{\T}$
in the neighborhood of $\bs^{*}$ by the first-order local polynomial,
\[
\widetilde{\psi}_{\lp}(\bs;\phi)=\left(\begin{array}{c}
\phi_{1}^{\T}d(\bs^{*}-\bs)\\
\vdots\\
\phi_{p}^{\T}d(\bs^{*}-\bs)
\end{array}\right)_{p\times1},\ \phi_{j}=\left(\begin{array}{c}
\phi_{j,0}\\
\phi_{j,1}\\
\phi_{j,2}
\end{array}\right)_{3\times1},\ d(\bs^{*}-\bs)=\left(\begin{array}{c}
1\\
s_{1}^{*}-s_{1}\\
s_{2}^{*}-s_{2}
\end{array}\right)_{3\times1},
\]
and $\phi=(\phi_{1,0},\ldots,\phi_{p,0},\phi_{1,1},\ldots,\phi_{p,1},\phi_{1,2},\ldots,\phi_{p,2})^{\T}$
is the vector of unknown coefficients. Although we use the first-order
local polynomial approximation, extensions to higher-order approximations
are straightforward with heavier notation. As established in Section
\ref{sec:Basic-setup}, $\psi^{*}(\bs^{*})$ is identified based on
the estimating function (\ref{eq:opt ef}), so we adopt the local
estimating equation approach (\citealp{carroll1998local}) with geographical
weighting. We propose a geographically weighted estimator $\widehat{\phi}_{\tau}(\bs^{*})$
by solving 
\begin{equation}
    \begin{split}
        &\mP_{n}\left[\omega_{\tau}(||\bs^{*}-\bs||)d(\bs^{*}-\bs)\otimes G\{V;\widetilde{\psi}_{\lp}(\bs;\phi,\mu, e)\}\right],\label{eq:GWE}\\
&=
\mP_n\left\{\omega_{\tau}(\|\boldsymbol{s}^*-\boldsymbol{s}\|)
\begin{pmatrix}
1\\
s_1^*-s_1\\
s_2^*-s_2
\end{pmatrix}\right.\\
&\left.\otimes 
\left(\sum_{m=1}^K\sum_{k=m}^K 
q_{k,m}(\overline{L}_m)
\left[
H_k^{(\overline{A}_{m-1})}\{\tilde{\psi}_{\rm lp}(\boldsymbol{s};\phi)\}
-\mu_{m,k}(\overline{L}_m)
\right]
\{A_m-e(\overline{L}_m)\}
\right)\right\}=0,
    \end{split}
\end{equation}
for $\phi$, where $M_{1}\otimes M_{2}$ denotes the Kronecker product
of $M_{1}$ and $M_{2}$, $H_k^{(\overline{A}_{m-1})}\{\tilde{\psi}_{\rm lp}(\boldsymbol{s};\phi)\}
=Y_k - \sum_{l=m}^k \gamma_{l,k}\{\tilde{\psi}_{\rm lp}(\boldsymbol{s};\phi)\}$, and $\omega_{\tau}(\cdot)$ is a spatial
kernel function with a scale parameter $\tau$. The first $p$-vector
$\widehat{\phi}_{\tau,0}(\bs^{*})$ in $\widehat{\phi}_{\tau}(\bs^{*})$
estimates $\psi^{*}(\bs^{*})$. The estimating equation (\ref{eq:GWE})
assigns more weight to observations nearby than those far from the
location $\bs^{*}.$ The commonly-used weight function is $\omega_{\tau}(||\bs^{*}-\bs||)=\tau^{-1}K\{||\bs^{*}-\bs||/\tau\}$,
where $K(\cdot)$ is the Gaussian kernel density function. The scale
parameter $\tau$ is the bandwidth determining the scale of spatial
treatment effect heterogeneity; $\psi^{*}(\bs)$ is smooth over $\bs$
when $\tau$ is large, and vice versa.

We illustrate the geographically weighted estimator of $\psi^{*}(\bs^{*})$
with a simple example, which allows an analytical form.

\begin{example}
\label{eg:linear-SNMM-weight}For the spatially varying structural nested mean
model in Example \ref{eg:linear-SNMM}, the geographically weighted
estimator of $\psi^{*}(\bs^{*})$ is the first $p$-vector of $\widehat{\phi}_{\tau}(\bs^{*})$ solving (\ref{eq:GWE}), with $\gamma_{m,k}\{\tilde{\psi}_{\rm lp}(\bs;\phi)\} = 
    \delta_{\rm lp}(\bs;\phi) 
    \exp \left[
    -\{t_k-t_m-\mu_{\rm lp}(\bs;\phi)\}^2/\{2\sigma_{\rm lp}(\bs;\phi)^2\}
    \right]a_m$, where
\begin{align*}
   &\delta_{\rm lp}(\boldsymbol{s};\phi) = \delta + \phi_{1,1}(s_1^*-s_1) + \phi_{1,2}(s_2^*-s_2),\\
   &\mu_{\rm lp}(\boldsymbol{s};\phi) = \mu + \phi_{2,1}(s_1^*-s_1) + \phi_{2,2}(s_2^*-s_2),\\
   &\sigma_{\rm lp}(\boldsymbol{s};\phi)^2 = \sigma^2 + \phi_{3,1}(s_1^*-s_1) + \phi_{3,2}(s_2^*-s_2),
\end{align*}
and $\phi = (\delta, \mu, \sigma^2, 
    \phi_{1,1}, \phi_{2,1}, \phi_{3,1},
     \phi_{1,2}, \phi_{2,2}, \phi_{3,2})$.
\end{example}


\begin{remark}It is worth discussing an alternative way to approximate
$\mu_{m,k}(\overline{L}_{m})$ by noticing that $\mu_{m,k}(\overline{L}_{m})=\E\{Y_{k}^{(\overline{A}_{m-1})}\mid\overline{L}_{m}\}$.
It amounts to identifying subjects who followed a treatment regime
$(\overline{A}_{m-1},\overline{0})$ and fitting the outcome mean model based
on $Y_{k}$ and $\overline{L}_{m}$ among these subjects.

\end{remark}

\subsection{Asymptotic properties}\label{subsec:asymptotic}

We show that the proposed estimator has an appealing double robustness
property in the sense that the consistency property requires that
one of the nuisance function models is correctly specified, not necessarily
both. This property adds protection against possible misspecification
of the nuisance models. Below, we establish the asymptotic properties
of $\widehat{\phi}_{\tau,0}(\bs^{*})$, including robustness, asymptotic
bias and variance. Assume the observed locations are continuously
distributed over a compact study region. Let $f_{\bs}(\bs)$ be the
marginal density of the observed locations $\bs$, which is bounded
away from zero. Also, assume that $\psi(\bs)$ is twice differentiable
with bounded derivatives. The kernel function $K(\cdot)$ is bounded
and symmetric, and the bandwidth satisfies $\tau\rightarrow0$ and $n\tau^{2}\rightarrow\infty$
as $n\rightarrow\infty$.

\begin{theorem}\label{thm:asymp}

Suppose Assumption \ref{assumpM-0}, the sequential randomization
assumption and the regularity conditions presented in the Supplementary
Material hold. Let $\widehat{\phi}_{\tau}(\bs^{*})$ be the solution
to (\ref{eq:GWE}) with $\mu_{m,k}(\overline{L}_{m})$ and $e(\overline{L}_{m})$
replaced by their estimates $\widehat{\mu}_{m,k}(\overline{L}_{m})$ and
$\widehat{e}(\overline{L}_{m})$. Let $\widehat{\phi}_{\tau,0}(\bs^{*})=e_{1}^{\T}\widehat{\phi}_{\tau}(\bs^{*}),$
where $e_{1}=(I_{p\times p},0_{p\times p},0_{p\times p})^{\T}$. If
either $\mu_{m,k}(\overline{L}_{m})$ or $e(\overline{L}_{m})$ is correctly
specified, $\widehat{\phi}_{\tau,0}(\bs^{*})$ is consistent for $\psi^{*}(\bs^{*})$,
and 
\[
\left\{ n\tau^{2}f_{\bs}(\bs^{*})\right\} ^{-1/2}\left\{ \widehat{\phi}_{\tau,0}(\bs^{*})-\psi^{*}(\bs^{*})-\tau^{2}\mathcal{G}_{b}(\bs^{*})\right\} \rightarrow\mathcal{N}\{0_{p\times1},\mathcal{G}_{v}(\bs^{*})\},
\]
where $\mathcal{G}_{b}\{\bs^{*},K,\psi(\bs^{*})\}$ and $\mathcal{G}_{v}\{\bs^{*},K,\psi(\bs^{*})\}$
do not depend on $\tau$.

\end{theorem}

The proof and the expressions of $\mathcal{G}_{b}\{\bs^{*},K,\psi(\bs^{*})\}$
and $\mathcal{G}_{v}\{\bs^{*},K,\psi(\bs^{*})\}$ are presented in
the Supplementary Material.

\subsection{Bandwidth selection using a balancing criterion \label{subsec:bandwidth}}

We use the K-fold cross-validation to select the bandwidth $\tau$.
An important question arises about the loss function. We propose a
new objective function using a balancing criterion. The key insight
is that with a good choice of $\tau,$ the mimicking potential outcome
$H_{k}^{(\overline{A}_{m-1})}\{\widehat{\phi}_{\tau,0}(\bs^{*})\}$ is
approximately uncorrelated to $A_{m}$ given $\overline{L}_{m}$; therefore,
the distribution of $H_{k}^{(\overline{A}_{m-1})}\{\widehat{\phi}_{\tau,0}(\bs^{*})\}$
is balanced between $A_{m}=1$ and $A_{m}=0$ among the group with
the same $\overline{L}_{m}$. If $\overline{L}_{m}$ contains continuous variables,
the balance measure is difficult to formulate because it involves
forming subgroups by collapsing observations with similar values of
$\overline{L}_{m}$. To avoid this issue, inspired by the estimating functions,
we formulate the loss function as 
\begin{multline}
G_{\loss}\{V;\psi(\bs^{*})\}=\left\vert \mP_{n}\omega_{\tau}(||\bs^{*}-\bs||)\vphantom{\sum_{k}^{K}}\right.\\
\left.\times\sum_{m=1}^{K}\sum_{k=m}^{K}\left[H_{k}^{(\overline{A}_{m-1})}\{\widehat{\phi}_{\tau,0}(\bs^{*})\}\right.\left.\vphantom{H_{k}^{(\overline{A})}}-\widehat{\mu}_{m,k}(\overline{L}_{m})\right]\{A_{m}-\widehat{e}(\overline{L}_{m})\}\right\vert .\label{eq:opt ef-1}
\end{multline}
If $\tau$ is too small, $\widehat{\phi}_{\tau,0}(\bs^{*})$ has a
large variance and also $\omega_{\tau}(||\bs^{*}-\bs||)$ is only
nontrivial for few locations in the $\tau$-neighborhood of $\bs^{*}$,
which leads to a large value of $G_{\loss}\{V;\psi(\bs^{*})\}$; while
if $\tau$ is too large, $\widehat{\phi}_{\tau,0}(\bs^{*})$ has a
large bias, which translates to a large loss too. Therefore, a good
choice of $\tau$ balances the trade-off between variance and bias.

\subsection{Bias correction}\label{subsec:bc}

Theorem \ref{thm:asymp} provides the asymptotic bias formula, which
however involves derivatives of $\psi^{*}(\bs^{*})$ and is difficult
to approximate. Following \citet{ruppert1997empirical}, we extend
the empirical bias correction method to the geographically
weighted framework. For a fixed location $\bs^{*},$ we calculate
$\widehat{\phi}_{\tau,0}(\bs^{*})$ at a series of $\tau$ over a
pre-specified range $\mathcal{T}=\{\tau_{1},\ldots,\tau_{L}\},$ where
$L$ is at least $3$. Based on Theorem \ref{thm:asymp}, the bias
function of $\widehat{\phi}_{\tau,0}(\bs^{*})$, with respect to $\tau,$
is of order $\tau^{2}$. This motivates a bias function of a form
$b(\tau;\nu)=\nu_{1}\tau^{2}+\cdots+\nu_{q}\tau^{q}$, where $q$ is an integer greater than $2$ and $\nu=(v_{1},\ldots,\nu_{q})^{\T}$
is a vector of unknown coefficients. Based on the pseudo data $\{\tau,\widehat{\phi}_{\tau,0}(\bs^{*}):\tau\in\mathcal{T}\}$,
fit a function $\E\{\widehat{\phi}_{\tau,0}(\bs^{*})\}=\nu_{0}+b(\tau;\nu)$
to obtain $\widehat{\nu}$. Then, we estimate the bias of $\widehat{\phi}_{\tau,0}(\bs^{*})$
by $b(\tau;\widehat{\nu}).$ The debiased estimator is $\widehat{\phi}_{\tau,0}^{\bc}(\bs^{*})=\widehat{\phi}_{\tau,0}(\bs^{*})-b(\tau;\widehat{\nu})$.

\subsection{Wild bootstrap inference\label{subsec:wild}}

For variance estimation, \citet{carroll1998local} proposed using
the sandwich formula in line with the Z-estimation literature. The
sandwich formula is justified based on asymptotics. To improve the
finite-sample performance, \citet{galindo2001bootstrap} proposed
a bootstrap inference procedure for local estimating equations. However,
this procedure involves constructing complicated residuals and a heuristic
modification factor. Thus, we suggest an easy-to-implement wild bootstrap
method for variance estimation of $\widehat{\phi}_{\tau,0}^{\bc}(\bs^{*})$.

For each bootstrap replicate, we generate exchangeable random weights
$\xi_{i}$ ($i=1,\ldots,n$ ) independent and identically distributed
from a distribution that has mean one, variance one and is independent
of the data; e.g., Exp$(1)$. The regular nonparametric bootstrap is included as a special case by adopting the multinomial distribution to generate the weights. Repeat the cross-validation for choosing
$\tau$, calculation of $\widehat{\phi}_{\tau,0}(\bs^{*})$, and bias-correction
steps but all steps are carried out using weighted analysis with $\xi_{i}$
for subject $i$. Importantly, we do not need to re-estimate the nuisance
functions for each bootstrap replication, because they converge faster
than the geographically weighted estimator. This feature can largely
reduce the computational burden in practice. The variance estimate
$\widehat{V}(\bs^{*})$ of $\widehat{\phi}_{\tau,0}^{\bc}(\bs^{*})$
is the empirical variance of a large number of bootstrap replicates.
With the variance estimate, we can construct the Wald-type confidence
interval as $\widehat{\phi}_{\tau,0}^{\bc}(\bs^{*})\pm z_{1-\alpha/2}\left\{ \widehat{V}(\bs^{*})\right\} ^{1/2}$.

\section{Extension to the settings with non-responses\label{sec:non-responses}}

In large longitudinal observational studies, non-responses are ubiquitous.
To accommodate non-responses, let $\overline{R}=(R_{0},\ldots,R_{K})$
be the vector of response indicators; i.e., $R_{m}=1$ if the subject
responded at $t_{m}$ and $0$ otherwise. With a slight abuse of notation,
let $V=(\overline{A},\overline{X},\overline{Y},\overline{R})$ be the full data. Let $\pi_{m}(V)=P(R_{m}=1\mid V)$
be the response probability at $t_{m}$. If the response probabilities
are known, an inverse probability weighted (IPW) estimating function 
\begin{multline}
G_{\mathrm{ipw}}\{V;\psi(\bs)\}=\sum_{m=1}^{K}\left\{ \sum_{k=m}^{K}\omega_{m:k}(\overline{R})q_{k,m}(\overline{L}_{m})\left[H_{k}^{(\overline{A}_{m-1})}\{\psi(\bs)\}-\mu_{m,k}(\overline{L}_{m})\right]\right\} \\
\times\{A_{m}-e(\overline{L}_{m})\},\label{eq:ipw}
\end{multline}
is unbiased at $\psi^{*}(\bs)$, where $\omega_{m:k}(\overline{R})=\prod_{l=m}^{k}\{R_{l}{\pi_{l}(V)}^{-1}\}$
is the inverse probability weights of responding from $t_{m}$ to
$t_{k}$. The geographically weighted local polynomial framework applies
by using (\ref{eq:ipw}) for estimating $\psi^{*}(\bs^{*})$.

In practice, $\pi_{m}(V)$ is unknown. We require further assumptions
for identification and estimation of $\pi_{m}(V)$. The most common
approach makes a missingness at random assumption \citep{rubin1976inference}
that $\pi_{m}(V)$ depends only on the observed data but not the missing
values. In smartphone applications with subject-initiated reporting,
whether subjects reported or not is likely to depend on their current
status. Then, the response mechanism depends on the possibly missing
values themselves, leading to an informative non-response mechanism.
In these settings, one can utilize a non-response instrument to help
identification and estimation of $\pi_{m}(V)$ (\citealp{wang2014instrumental,li2020robust}).
For illustration, we assume a simple informative response mechanism
that $\pi_{m}(V)=\pi(V_{m})$, where $V_{m}=(A_{m},X_{m},Y_{m},\overline{R}_{m-1})$;
i.e., $\pi_{m}(V)$ depends only on the current (possibly missing)
status $(A_{m},X_{m},Y_{m})$ and the number of historical responses.
Extension to more complicated mechanisms is possible at the expense
of heavier notation. We then posit a parametric response model, denoted
by $\pi_{m}(V_{m};\eta^{*})$ with an unknown parameter $\eta^{*}\in\mathcal{R}^{d}$.
For informative non-responses, the parameter $\eta^{*}$ is not identifiable
even with a parametric model \citep{wang2014instrumental}. We assume
that there exists an auxiliary variable $Z_{m}$ called a non-response
instrument that can be excluded from the non-response probability,
but are associated with the current status even when other covariates
are conditioned (see Condition C1 in Theorem 1 of \citealp{wang2014instrumental}
for the formal definition). Existence of such non-response instruments
depends on the study context and available data. For example, in the
Smoke Sense initiative, a valid non-response instrument is the smoke
plume data HMS measured at monitors, which is related to the subject's
variables but is unrelated to whether the subject reporting or not
after controlling for subject's own perceptions about risk.

With a valid instrument, $\eta^{*}$ is identifiable. Following \citet{robins1997analysis},
$\widehat{\eta}$ can be obtained by solving
\begin{equation}
\mP_{n}\sum_{m=0}^{K}\left\{ \frac{R_{m}}{\pi_{m}(V_{m};\eta)}-1\right\} h(Z_{m},\overline{R}_{m-1})=0,\label{eq:GMM}
\end{equation}
where $h(Z_{m},\overline{R}_{m-1})\in\mathbb{\mathcal{R}}^{d}$ is a function
of $(Z_{m},\overline{R}_{m-1})$. An additional complication involves estimating
the nuisance functions $\mu_{m,k}(\overline{L}_{m})$ and $e(\overline{L}_{m})$
in the presence of non-response, which now requires weighting similar
to that in (\ref{eq:ipw}). 
The IPW approach requires an accurate model for response probability. Following \cite{yang2022semiparametric} and \cite{coulombe2024quadruply}, future work could enhance robustness against model misspecification. We leave this for future research.
We summarize the stepwise procedures to obtain the proposed estimator under informative missingness in Algorithm \ref{alg:local_SNMM_kernel_weighting}, and defer the technical details to the Supplementary Material.
 \begin{algorithm}[!ht]
\caption{\label{alg:local_SNMM_kernel_weighting} Geographically weighted local polynomial estimation for structural nested mean models}
\textbf{Input}: 
Subjects' full records $V=({A}_m,{X}_m,{Y}_m)_{m=0}^{K}$, their spatial locations $\boldsymbol{s}=(s_1, s_2)$, the grid for selecting $\tau$, the polynomial order $q$ for modeling the bias function, and the bootstrap size $B$.\\
\textbf{Step 1.} Fit a response probability model $\widehat{\pi}_m(V)$ via (\ref{eq:GMM}).\\
\textbf{Step 2.} Fit a propensity score model $\widehat{e}(\overline{L}_{m})$ by solving
\begin{equation}
\mP_{n}\left[
\sum_{m=0}^K 
R_m \widehat{\pi}^{-1}_m(V)
\left\{
A_m e^{-1}(\overline{L}_m) - 1
\right\} h(\overline{L}_m)
\right]=0.\label{eq:GWLP0-ps}
\end{equation}
\\
\textbf{Step 3.} For each location $\bs^{*}$, obtain a
initial estimator $\widehat{\phi}_{\tau}^{[0]}(\bs^{*})$ by solving (\ref{eq:GWE}) with $G\{V;\widetilde{\psi}_{\lp}(\bs;\phi)\}$ replaced by $G_{\ipw}\{V;\widetilde{\psi}_{\lp}(\bs;\phi)\}$ in (\ref{eq:ipw}), where ${\phi}_{\tau}^{[0]}(\bs^{*})$ is the first $p$-vector in $\widetilde{\psi}_{\lp}(\bs;\phi)$.\\
\textbf{Step 4.} Using the pseudo outcome $H_{k}^{(\overline{A}_{m-1})}\left\{ \widehat{\phi}_{\tau,0}^{[0]}(\bs^{*})\right\} $
and $\overline{L}_{m}$, fit an outcome mean model $\widehat{\mu}_{m,k}(\overline{L}_{m})$ using the IPW estimating equation. \\
\textbf{Step 5.} Obtain $\widehat{\phi}_{\tau,0}(\bs^{*})$ by solving the IPW version of (\ref{eq:GWE}) with $\widehat{e}(\overline{L}_{m})$ and $\widehat{\mu}_{m,k}(\overline{L}_{m})$.\\ 
\textbf{Step 6.} The debiased estimator $\widehat{\phi}_{\tau,0}^{\bc}(\bs^{*})=\widehat{\phi}_{\tau,0}(\bs^{*})-b(\tau;\widehat{\nu})$ with $\widehat{\nu}$ fitted by polynomial regression with order $q$.\\
Repeat Steps 1-6 for $B$ times with random weights $\xi_i (i=1,\cdots,n)$ to compute $\mP_n$, and obtain the empirical variance $\widehat{V}(\boldsymbol{s}^*)$ of the bootstrap replicates.\\
\textbf{Output}: A debiased local geographical kernel weighting estimator $\widehat{\phi}_{\tau,0}^{\bc}(\bs^{*})$ and its confidence interval $\widehat{\phi}_{\tau,0}^{\bc}(\bs^{*})\pm z_{1-\alpha/2}\left\{ \widehat{V}(\bs^{*})\right\} ^{1/2}$.
\end{algorithm}

\section{Simulation study\label{sec:simulation}}

We evaluate the finite-sample performance of the proposed estimator
on simulated datasets to evaluate the double robustness property.
We first consider the simpler case without non-responses in Section
\ref{subsec:without-non-response} and then consider the case with
non-responses in Section \ref{subsec:With-informative-nonresponse}
to mimic the Smoke Sense data. All codes are parallelized and executed on a computer with Intel(R) Core(TM) i7-8565U CPU, 16 RAM computer. The proposed geographically weighted estimation takes approximately $8$ hours when evaluated on a $10\times 10$ grid of spatial locations for one replicated dataset in our simulation study.

\subsection{Complete data without non-responses\label{subsec:without-non-response}}

 In this section, we simulate
$500$ datasets without non-response. In each dataset, we generate
$n$ locations. For each location, each subject are associated with $P$ covariates and the subject's $p$-th covariate process over $K=25$ weeks follows a Gaussian process with mean $0$, variance $0.5$ and a first-order
autocorrelation structure with lag-1 correlation $0.5$ for $p=1,\cdots,P$. We generate
the potential outcome process as $Y_{k}^{(\overline{0})}=
\sum_{p=1}^P X_{p,k}+\epsilon_{k}$,
for $k=1,\ldots,K$, where $X_{p,k}$ is the subject's $p$-th covariate at time $t_k$, and $\overline{\epsilon}$ follows a Gaussian process
with mean $0$, variance $0.25$, and a first-order autocorrelation
structure with lag-1 correlation $0.25$. We generate the treatment
process as $\overline{A}$, where $A_{k}\sim$ Binomial\{$e(\overline{L}_{k})$\}
with logit$\{e(\overline{L}_{k})\}=-1+0.5\sum_{p=1}^P X_{p,k}+0.25$cum$(\overline{A}_{k-1})$ and cum$(\overline{A}_{k-1})=\sum_{m=0}^{k-1}A_{m}$. The observed outcome process is $Y_{k}=Y_{k}^{(\overline{0})}+\sum_{m=0}^{k-1}\gamma_{m,k}\{\psi^{*}(\bs)\}$, where $\gamma_{m,k}\{\psi^{*}(\bs)\}=\psi^{*}(\bs)A_{m},$ if $k=m+1$ and zero otherwise. We consider three treatment effect specifications for location $\bs=(s_{1},s_{2})$. In the first two scenarios, $n=1076$ locations are uniformly sampled over a unit square with $P=1$. In the third scenario, we consider a similar dimension of covariates as the Smoke Sense data with $P=25$ and an unobserved spatial covariate $Z$ in an irregular grid. For a given location $\boldsymbol{s}^*=(s_1^*, s_2^*)$, we generate another covariate process $\overline{Z}_K=(Z_1,\cdots,Z_K)^{\intercal}$ from a Gaussian process with mean $\mu_Z(s_1,s_2)=\sqrt{s_1^2+s_2^2}/100$, variance 0.5 and a first-order autocorrelation structure with lag-1 correlation 0.5; and the treatment effect $\gamma_{m,k}\{\psi^*(\bs)\} = Z_m A_m$. In the fitting process, $\{Z_m, m=1,\cdots,K\}$ are merely a set of unobserved realizations of random variables. Thus, the true treatment effect $\psi^*(\bs)$ should be $\mu_Z(s_1,s_2)$, which is the mean of $\overline{Z}_K$. Mimicking the spatial dependence in the Smoke Sense data, the data points with size $n=1882$ are selected spread over an irregular grid.

\begin{enumerate}
\item[(S1)] $\psi^{*}(\bs)=\exp(s_{1}+s_{2})$; 
\item[(S2)] $\psi^{*}(\bs)=\sin\{2(s_{1}+2s_{2}-1)\}$;
\item[(S3)] $\psi^*(\bs) = \mu_Z(s_1,s_2)$.
\end{enumerate}

We then consider estimating $\psi^{*}(\bs^{*})$ at four locations. For (S1) and (S2), $\bs^{*}=(s_{1}^{*},s_{2}^{*})$ and $s_{j}^{*}\in\{0.25,0.75\}$ ($j=1,2$). For (S3), the evaluation locations are $\bs_1^*=(37.4 -122.3)$, $\bs_2^*=(39.2, -122.3)$, $\bs_3^*=(37.4, -120.2)$, and $\bs_4^*=(39.2, -120.2)$ whose lattidute and longitude degrees are $1/5$ and $4/5$ quantiles of the lattidute and longitude degrees in the real data set.

To investigate the double robustness in Theorem \ref{thm:asymp},
we consider two models for $\mu_{m,k}(\overline{L}_{m})$: (a) a correctly
specified linear regression model and (b) a misspecified model by
setting $\widehat{\mu}_{m,k}(\overline{L}_{m})=0$. We also consider two
models for $e(\overline{L}_{m})$: (a) a correctly specified logistic regression
model with predictors $X_{m}$ and cum$(\overline{A}_{m-1})$ and (b) a
misspecified logistic regression model with predictors $X_{m}^{2}$
and cum$(\overline{A}_{m-1})^{2}$. For all estimators, we consider a grid
of geometrically spaced values for $\tau$ from $\{\exp(0.005),\exp(0.05)\}$
and use 5-fold cross validation in Section 4.4
to choose $\tau$; see the Supplementary Material for the Monte Carlo average of the selected $\tau$ over a grid of spatial locations and additional simulations. The integer $q$ in the bias function is chosen to be $3$. We use the wild bootstrap for variance
estimation with the bootstrap size $200$.

Table \ref{tab:res1} reports the simulation results for Scenarios
(S1)--(S3), respectively. When either the model for the propensity
score or the model for the outcome mean is correctly specified, the
proposed estimator has small bias for all treatment effects at all
locations across three scenarios. These results confirm the double
robustness in Theorem \ref{thm:asymp}. Moreover, under these cases,
the wild bootstrap provides variance estimates that are close to the
true variances and good coverage rates that are close to the nominal
level.

\begin{table}
\caption{\label{tab:res1}Simulation results for complete data. The Monte Carlo average ($\times10^{-2},$ Est) and variance ($\times10^{-3}$, Var) of the estimators, variance estimators ($\times10^{-3}$, Ve), and coverage rate ($\%$, Cr) of $95\%$ confidence intervals. The three scenarios (S1)-(S3) are different values of the true spatially varying effect, and the results are reported separately for four fixed locations ($\bs_{1}^{*}$-$\bs_{4}^{*}$). The methods vary by whether they have the correct model for the outcome mean $\mu_{m,k}(\overline{L}_{m})$ and the propensity score $e({\overline{L}}_{m})$.}
\centering
\resizebox{\textwidth}{!}{%
\vspace{0.25cm}

\begin{tabular}{clcccccccccccccc}
\toprule 
 &  & \multicolumn{4}{c}{Scenario (S1)} &  & \multicolumn{4}{c}{Scenario (S2)} &  & \multicolumn{4}{c}{Scenario (S3)}\tabularnewline
\midrule 
 &  & $\bs_{1}^{*}$  & $\bs_{2}^{*}$  & $\bs_{3}^{*}$  & $\bs_{4}^{*}$  &  & $\bs_{1}^{*}$  & $\bs_{2}^{*}$  & $\bs_{3}^{*}$  & $\bs_{4}^{*}$  &  & $\bs_{1}^{*}$  & $\bs_{2}^{*}$  & $\bs_{3}^{*}$  & $\bs_{4}$\tabularnewline
 & True & 164.9  & 271.8  & 271.8  & 448.2  &  & -47.9  & 47.9  & 99.7  & 59.80 & & 127.9 & 128.4 & 125.9 & 126.5 \tabularnewline
\midrule 
\multicolumn{16}{c}{Propensity score model $(\checked)$}\tabularnewline
\midrule 
 & Est   & 165.8  & 270.0  & 270.4  & 449.6  &  & -47.4  & 49.0  & 100.8  & 57.2 & & 127.8 & 128.9 & 125.9 & 126.7\tabularnewline
Mean  & Var  & 18.9  & 14.6  & 17.7  & 18.0  &  & 17.0  & 13.7  & 16.9  & 18.1 & & 1.0 & 1.5 & 2.4 & 2.3\tabularnewline
model $(\checked)$  & Ve   & 16.9  & 16.5  & 16.8  & 18.5  &  & 17.1  & 17.0  & 18.0  & 19.4 &  & 0.8 & 2.8 & 1.3 & 1.9 \tabularnewline
 & Cr  & 93.2  & 96.2  & 94.6  & 94.8  &  & 93.4  & 95.8  & 95.0  & 93.8 & & 95.1 & 95.2 & 94.0 & 94.6\tabularnewline
\midrule 
 & Est  & 167.4  & 270.7  & 270.6  & 451.8  &  & -45.8  & 49.9  & 100.6  & 59.3 & & 128.6&125.9 &123.3&125.0\tabularnewline
Mean  & Var   & 36.3  & 33.8  & 39.0  & 35.2  &  & 36.9  & 33.0  & 37.5  & 37.6 &  & 12.3 & 40.7 & 20.1 & 38.7\tabularnewline
model $(\times)$  & Ve   & 37.9  & 36.0  & 37.0  & 36.3  &  & 37.9  & 36.4  & 37.4  & 38.1 &  & 16.7 & 56.6 & 28.0 & 38.7\tabularnewline
 & Cr  & 93.6  & 96.2  & 95.0  & 94.4  &  & 94.2  & 95.4  & 94.8  & 95.40 & & 95.3 & 95.6 & 92.4 & 97.6\tabularnewline
\midrule 
\multicolumn{16}{c}{Propensity score model $(\times)$}\tabularnewline
\midrule 
 & Est  & 166.0  & 269.9  & 270.3  & 449.9  &  & -47.1  & 49.0  & 100.3  & 57.4 & & 128.1 & 128.4 & 126.0 & 127.1\tabularnewline
Mean  & Var   & 19.6  & 16.3  & 17.2  & 19.7  &  & 20.8  & 16.7  & 16.9  & 21.1 & & 1.1 & 4.1 & 2.1 & 2.5\tabularnewline
model $(\checked)$  & Ve   & 18.8  & 17.8  & 19.0  & 20.1  &  & 18.9  & 18.1  & 19.7  & 20.8 &  & 1.2 & 4.3 & 2.1 & 2.9 \tabularnewline
 & Cr & 93.6  & 95.8  & 94.6  & 93.8  &  & 92.4  & 95.2  & 95.6  & 94.8 &  & 95.8 & 95.6 & 93.6 & 96.8 \tabularnewline
\midrule 
 & Est  & 213.5  & 317.9  & 317.4  & 498.1  &  & 0.1  & 96.9  & 147.7  & 106. & & 1467 & 1465 & 1472 & 1470\tabularnewline
Mean  & Var  & 38.0  & 36.3  & 48.9  & 33.6  &  & 33.9  & 35.7  & 48.7  & 36.0 &  & 244 & 657 & 469 & 551\tabularnewline
model $(\times)$  & Ve & 37.6  & 36.3  & 36.8  & 37.9  &  & 38.0  & 38.9  & 38.3  & 40.3 &  & 261 & 896 & 447 & 642 \tabularnewline
 & Cr   & 22.0  & 25.0  & 23.6  & 24.6  &  & 22.6  & 19.4  & 22.4  & 29.0 & & 0.0 & 0.0 & 0.0 & 0.0 \tabularnewline
\bottomrule
\multicolumn{16}{c}{{\normalsize $\checked$ (is correctly specified), $\times$ (is misspecified)}}
\end{tabular}}
\end{table}

\subsection{Informative non-responses\label{subsec:With-informative-nonresponse}}

In this section, we generate data with informative non-responses.
The data-generating processes are the same as in Section \ref{subsec:without-non-response},
except that $P=1$ for simplicity and a vector of response indicators
$\overline{R}$ is generated, where $R_{k}\sim$ Binomial\{$\pi_{k}(Y_{k},\overline{R}_{k-1})$\}
with logit$\{\pi_{k}(Y_{k},\overline{R}_{k-1})\}=-c+0.5Y_{k}+0.25$cum$(\overline{R}_{k-1})$, where $c=-1$ for (S1), (S2) and $c=-2$ for (S3).
If $R_{k}=1$, $(A_{k},{X}_{k},Y_{k})$ is observed, otherwise only
${X}_{k}$ is observed. Thus, ${X}_{k}$ is a non-response instrument.
We compare the following estimators. 
\begin{description}
\item [{GWLPc1:}] the geographically weighted local polynomial estimator
with the non-response weights setting to a constant $1$, which corresponds
to the proposed estimator with a misspecified model for the non-response
probability; 
\item [{GWLPipw:}] the proposed geographically weighted local polynomial
estimator with inverse probability of non-response weighting adjustment. 
\end{description}
For both estimators, we consider correctly specified models for the
outcome mean and propensity score. We use the same cross-validation
and wild bootstrap procedures as in Section \ref{sec:simulation}.

Table \ref{tab:res1-missing} reports the simulation results with
informative non-responses for Scenarios (S1)--(S3). The GWLPc1 estimator
without the inverse probability of non-response weighting adjustment
is biased, and its coverage rate is off the nominal level, suggesting
that informative missingness would lead to biased conclusions. The
GWLPipw estimator with proper weighting adjustment has small bias
for all locations across all scenarios.

\begin{table} 
\caption{\label{tab:res1-missing}Simulation results for data with informative non-responses. The Monte Carlo average ($\times10^{-2},$ Est) and variance ($\times10^{-3}$, Var) of the estimators, variance estimators ($\times10^{-3}$, Ve), and coverage rate ($\%$, Cr) of $95\%$ confidence intervals. The three scenarios (S1)-(S3) are different values of the true spatially varying effect, and the results are reported separately for four fixed locations ($\bs_{1}^{*}$-$\bs_{4}^{*}$). The methods vary by whether they have the correct (GWLPipw) or incorrect (GWLPc1) model for the missing data mechanism ($\pi_{m})$.}
\centering
\resizebox{\textwidth}{!}{%
\vspace{0.25cm}

\begin{tabular}{clcccccccccccccc}
\toprule 
 &  & \multicolumn{4}{c}{Scenario (S1)} &  & \multicolumn{4}{c}{Scenario (S2)} &  & \multicolumn{4}{c}{Scenario (S3)}\tabularnewline
\midrule 
 &  & $\bs_{1}^{*}$  & $\bs_{2}^{*}$  & $\bs_{3}^{*}$  & $\bs_{4}^{*}$  &  & $\bs_{1}^{*}$  & $\bs_{2}^{*}$  & $\bs_{3}^{*}$  & $\bs_{4}^{*}$  &  & $\bs_{1}^{*}$  & $\bs_{2}^{*}$  & $\bs_{3}^{*}$  & $\bs_{4}^{*}$\tabularnewline
 & True   & 164.9  & 271.8  & 271.8  & 448.2  &  & -47.9  & 47.9  & 99.7  & 59.8 &  & 127.9 & 128.5 & 125.9 & 126.5 \tabularnewline
\midrule 
 & Est  & 163.8  & 271.0  & 270.4  & 450.0  &  & -42.9  & 52.0  & 101.3  & 59.0 & & 127.7 & 132.6 & 126.9 & 128.5 \tabularnewline
Missing data  & Var  & 63.7  & 43.2  & 40.4  & 44.0  &  & 148.3  & 112.4  & 71.2  & 91.1 & & 36.6 & 182.0 & 46.0 & 75.5 \tabularnewline
model $(\checked)$  & Ve & 60.5  & 46.1  & 46.0  & 40.5  &  & 210.3  & 123.0  & 115.5  & 100.3 & & 33.5 & 167.5 & 44.5 & 78.8\tabularnewline
 & Cr  & 93.4  & 95.4  & 94.6  & 92.8  &  & 92.2  & 93.6  & 94.2  & 91.8 & & 93.6 & 94.4 & 94.4 & 93.3\tabularnewline
\midrule 
 & Est & 158.2  & 260.2  & 260.2  & 434.0  &  & -45.4  & 41.1  & 87.1  & 48.1 & & 151.6 & 148.4 & 148.3 & 148.6\tabularnewline
Missing data  & Var & 44.5  & 27.9  & 33.8  & 29.3  &  & 57.8  & 26.6  & 24.4  & 31.9 & & 13.2 & 81.1 & 21.0 & 35.7 \tabularnewline
model $(\times)$  & Ve  & 57.6  & 29.6  & 29.7  & 26.2  &  & 70.8  & 31.5  & 28.3  & 44.4 & & 13.6 & 79.7 & 20.8 & 36.3\tabularnewline
 & Cr  & 91.4  & 88.4  & 86.8  & 83.0  &  & 93.6  & 93.0  & 78.0  & 85.6 & & 46.6 & 86.8 & 63.7 & 75.0 \tabularnewline
\bottomrule
\multicolumn{16}{c}{{\normalsize$ \checked$ (is correctly specified), $\times$ (is misspecified)}}
\end{tabular}
}
\end{table}

\section{Smoke Sense data application \label{sec:Real data}}

We now apply the new methodology to the Smoke Sense data to estimate
heterogeneous effects of protective measures to mitigate the {health} impact
of wildland fire smoke. We follow the basic setup in Section \ref{sec:Basic-setup},
where the data are recorded monthly ( $t_{k}=k$ is the $k$th month
after registration).

To model the treatment effect, a complication arises because taking
protective behaviors might not have immediate effects on health outcomes
and also the effect on the future outcomes eventually reduces to zero
for large time lags. Taking these features into account, we consider
the treatment effect model of a squared exponential function of the
time lag $(t_{k}-t_{m})$ as 
\begin{equation}
\gamma_{m,k}(\psi^{*})=\delta\exp\left\{ -\frac{(t_{k}-t_{m}-\mu)^{2}}{2\sigma^{2}}\right\} a_{m},\label{eq:GaussianM}
\end{equation}
where $\psi^{*}=(\delta,\mu,\sigma^{2})$ is the vector of parameters. A negative sign of $\delta$
indicates that the treatment is beneficial in reducing the number
of adverse health outcomes, a larger magnitude of $\delta$ means a
larger maximum treatment effect, vice versa, and $\sigma^2$ determines
the duration of the effect across time lags. We consider fitting a
global model (\ref{eq:GaussianM}) in Section \ref{subsec:Global-estimation}
and a local model with spatially varying $\psi^{*}(\bs)$ in Sections
\ref{subsec:Local-constant-estimation} and \ref{subsec:analysis_cities}. To assess the fitted nuisance functions and the local treatment effect models, a set of diagnostic analyses are conducted in Section S6.2 of the Supplementary Material. According to the model diagnostic statistics, there is no significant evidence to reject the conditional independence between $Y_k 
- \sum_{l=m}^k 
\gamma_{l,k}\{\widehat{\psi}^*(\boldsymbol{s}^*)\}-\widehat{\mu}_{m,k}(\overline{L}_m)$ and $A_m-\widehat{e}(\overline{L}_m)$ for most selected locations and cities. This suggests that $\overline{L}_m$ adequately captures the confounding variables, and $Y_k - \sum_{l=m}^k \gamma_{l,k}\{\widehat{\psi}^*(\boldsymbol{s}^*)\}$ closely resembles the potential outcome $Y_k^{(\overline{a}_{m-1})}$. Thus, we conclude that the models $\gamma_{m,k}\{\widehat{\psi}^*(\boldsymbol{s}^*)\}$, $\widehat{\mu}_{m,k}(\overline{L}_m)$ and $\widehat{e}(\overline{L}_m)$ are well-fitted.

For the missing values of the covariates, we impute the missing time-varying
variables by carrying forward the last observations, and the missing
time-independent variables by mean imputation, i.e., the average and
max frequency values to impute the continuous/ordinal variables and
categorical variables, respectively. The reason that we employ this imputation strategy is to preserve the assumption of sequential ignorability by avoiding the imputation of missing values based on future observations. Furthermore, our proposed method is evaluated after utilizing multiple imputations to fill in the missing covariates in Section S6.4 of the Supplementary Materials, where the global treatment effects exhibit a similar pattern. It should be noted that the randomness due to the missing covariates is accounted for by re-imputing each bootstrapped dataset in the wild bootstrap procedure.

We adopt the non-response instrument
variable approach in Section \ref{sec:non-responses} to adjust for
informative non-responses. We assume the response probability $\pi_{m}(V_{m})$
follows a logistic regression with a linear predictor $\eta^{\T}\{\begin{array}{cccc}
1 & A_{m} & Y_{m} & {\rm cum}(\overline{R}_{m-1})\end{array}\}$. To identify and estimate $\eta$, we use the non-response instrument
${\rm HMS}_{m}$, the true smoke status, which affects the subject's
health status but is unrelated to whether the subject reporting or
not after controlling for the subject's own perceptions about risk.
The estimator of $\eta$ is obtained by solving the estimating equation
(\ref{eq:GMM}) with $h(Z_{m},\overline{R}_{m-1})=\{\begin{array}{cccc}
1 & {\rm cum}(\overline{R}_{m-1}) & \text{HMS}_{m} & \text{HMS}_{m}^{2}\end{array}\}^{\T}$. The solution is $\hat{\eta}=(0.04,0.05,0.11,0.14)^{\T}$, indicating
that more active participants with worse health outcomes are more
likely to report. The proposed estimation (\ref{eq:ipw}) requires approximately $10$ hours to be completed for a single spatial location of the Smoke Sense data under the treatment effect model (\ref{eq:GaussianM}).

\subsection{Global estimation\label{subsec:Global-estimation}}

We first consider fitting a global model and estimate the spatially
stationary parameters $\delta,\sigma^{2}$ and $\mu$ by solving the
estimating equation \eqref{eq:opt ef}, which yields estimates $(\widehat{\delta},\hat{\mu},\hat{\sigma}^{2})=(-0.3,7.0,3.0)$. Besides considering the Gaussian distributional-based treatment effect model as \eqref{eq:GaussianM}, for sensitivity analysis, we also tried the Gamma distributional-based model as shown in the Supplementary Materials, which gives a similar pattern.

\begin{figure}[ht!]
\begin{center}
\includegraphics[width=.75\linewidth]{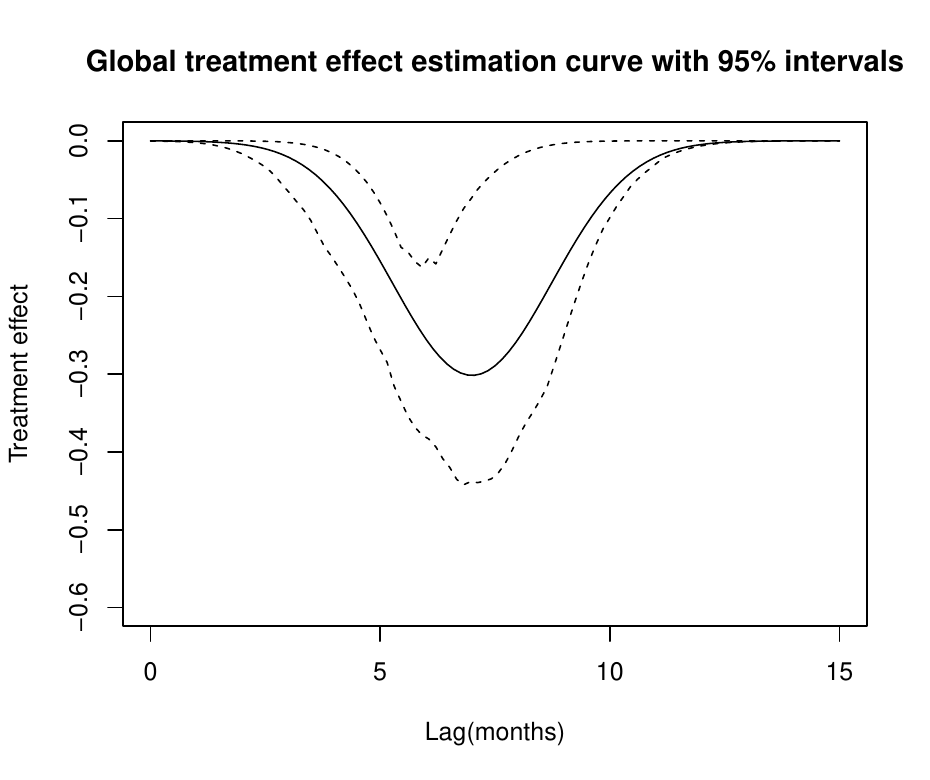} 
\par\end{center}
\centering{}\caption{Global treatment effect curve by time lag. The lag ($k-m$) is the
time between treatment (at the $m$th month) and outcome (at the $k$th
month). The solid line is the treatment effect $\gamma_{m,k}(\hat{\psi}^{*})$
evaluated at the estimated values $\hat{\psi}^{*}=(\widehat{\delta},\hat{\mu},\hat{\sigma}^{2})=(-0.3,7.0,3.0)$.
The dashed lines are the $95\%$ confidence bands constructed from
$200$ bootstrapped samples. \label{fig:Global-treatment-effect}}
\end{figure}

Figure \ref{fig:Global-treatment-effect} shows the estimated global
treatment effect curve with peak height ($\widehat{\delta}=-0.3$ symptoms),
peak location ($\hat{\mu}=7.0$ month), and duration ($\approx1$
year). The results suggest a long-lasting effect of taking protective
measures in reducing adverse health outcomes. The estimated lag between
taking protective measures and the maximum reduction in symptoms is
$7.0$ months and the estimated reduction in symptoms at this lag
is $0.3$ symptoms. With the wild bootstrap, the $95\%$ confidence
interval for $\delta$ is $(-0.5,-0.2)$, and thus the treatment effect is statistically
significant. We also conduct a sensitivity analysis with a different $\gamma_{m,k}(\psi^*)$. The results in Section S4.1 in the supplementary material suggest that the conclusions are robust to the functional form of  $\gamma_{m,k}(\psi^*)$. 

\subsection{Local estimation\label{subsec:Local-constant-estimation}}

\begin{figure}[ht!]
\begin{center}
\includegraphics[width=\textwidth]{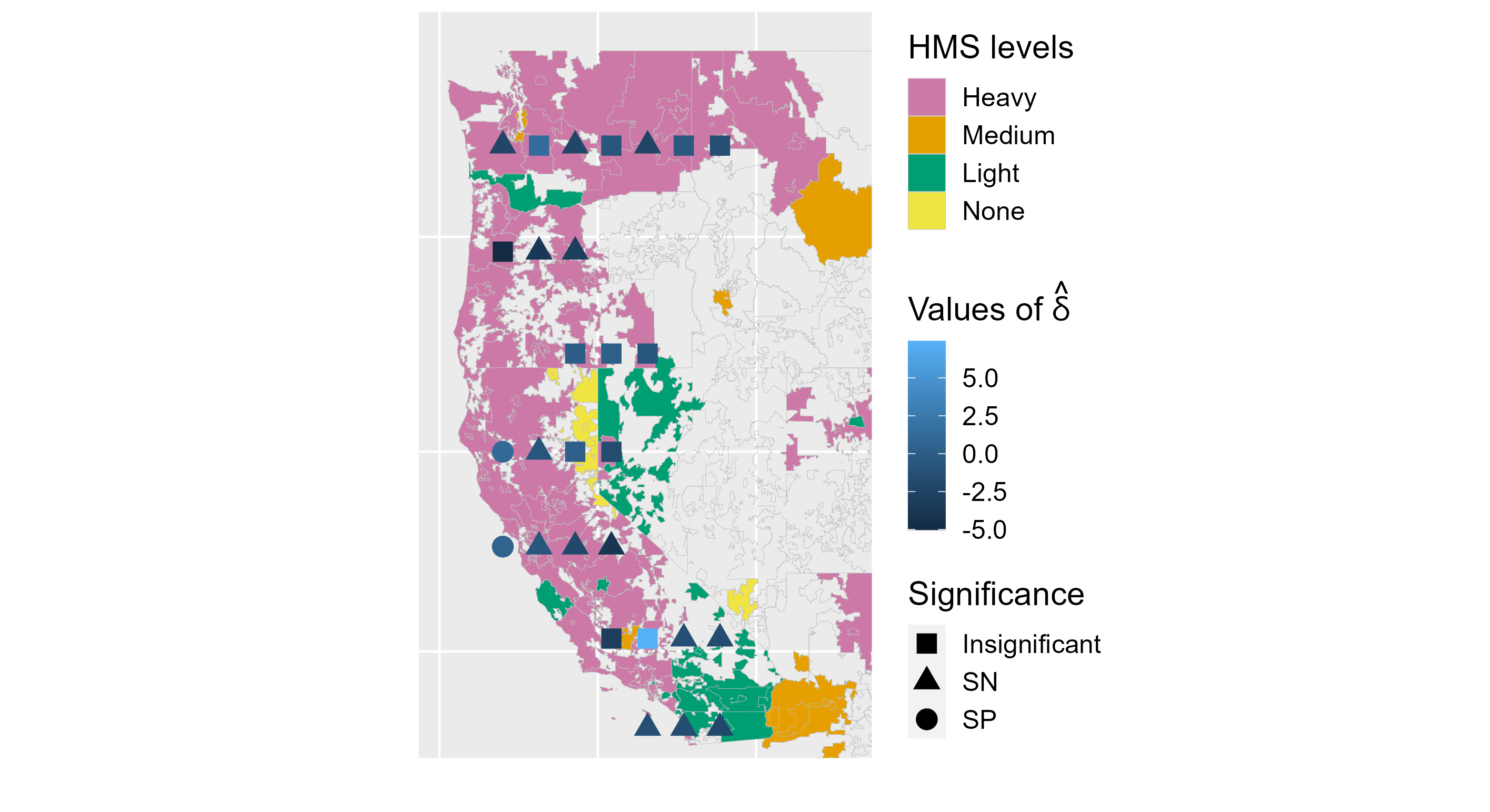} 
\par\end{center}
\caption{Estimated treatment effect $\widehat{\delta}$ for the $28$ locations in the
Western US. The shapes of symbols denote whether $\widehat{\delta}$ is
insignificant, significantly negative (SN) or significantly positive
(SN). The background color is the maximum daily HMS smoke density
in each three-digit zip code. The units of effect is the number of
symptoms. \label{fig:psi_bc}}
\end{figure}

The global model assumes the treatment effect curve is the same at
all locations. Figure \ref{fig:map_num_users} shows that the participants
of the Smoke Sense Initiative spread all over the west coast of the
US. It is likely that the treatment effect can have spatial heterogeneity
due to the large and diverse socially- and environmentally-diverse
study domain that is not explained by the observed covariates. In order to investigate the spatially varying treatment
effect, we fit a local model using geographical weighting \eqref{eq:GWE}.
To reduce the computational burden, we use local constant approximation
instead of local linear approximation. We choose spatial locations
for evaluation as follows. We first set a grid of $56$ locations
over the west coast of the US; the locations are combinations of 7
equally-spaced values for the latitude coordinate from $33$ to $47$
and $8$ equally-spaced values for the longitude coordinates from
$-123$ to $-115$. Among these locations, we select $28$ locations
that have samples within the one-degree neighborhood and have
more than $0.5\%$ samples within the two-degree neighborhood. Table S4 
in the Supplementary Material reports the estimates $\widehat{\delta},\hat{\mu}$
and $\hat{\sigma}^{2}$ and their confidence intervals at the $28$ locations,
and Figure \ref{fig:psi_bc} shows the color map of $\widehat{\delta}$. One important finding is that users located in the southwest are more likely to experience significant beneficial treatment effects.

\begin{figure}
    \centering
    \includegraphics[width=0.9\textwidth]{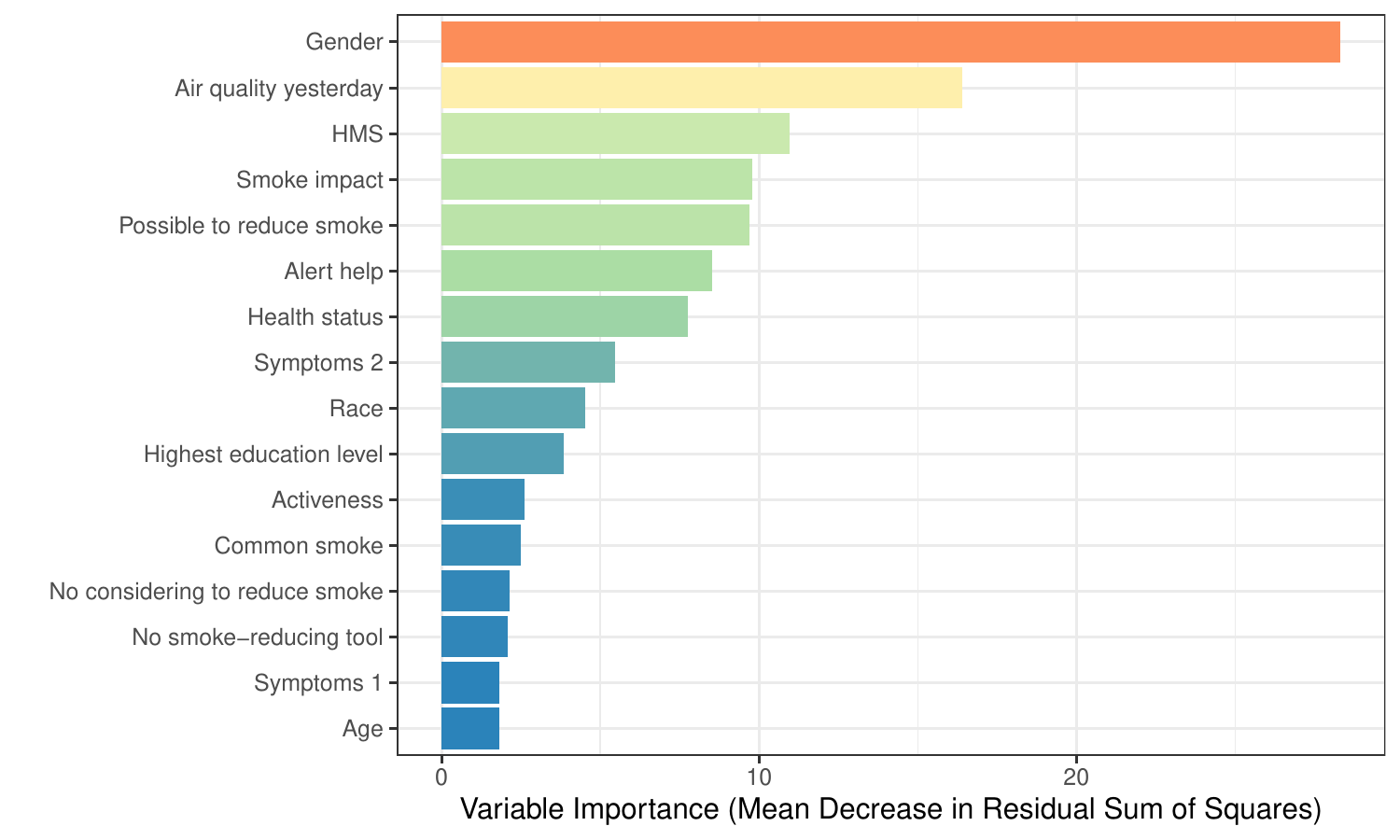}
    \caption{Variable importance plot of the random forest regression model in terms of the mean decrease in residual sum of squares.}
    \label{fig:imp_rf}
\end{figure}

To gain insights into the primary covariates driving the spatial variation in the estimated treatment effect, we consider fitting a random forest model for the estimated treatment effects $\widehat{\delta}(\boldsymbol{s})$ against all the baseline characteristics and two time-varying environmental variables ("Air quality yesterday" and "HMS"). The variable importance, depicted in Figure \ref{fig:imp_rf}, is arranged based on their impacts in reducing the residual sum of squares. Upon analyzing Figure \ref{fig:imp_rf}, we observe that "Gender" and "Air quality yesterday" emerge as the two most critical covariates responsible for the spatial variation in the treatment effects. Hence, these two variables play a significant role in explaining the spatial-specific treatment effects.

\subsection{Analysis for cities and national parks\label{subsec:analysis_cities}}

\begin{figure}[ht!]
\begin{center}
\includegraphics[width=0.4\textwidth]{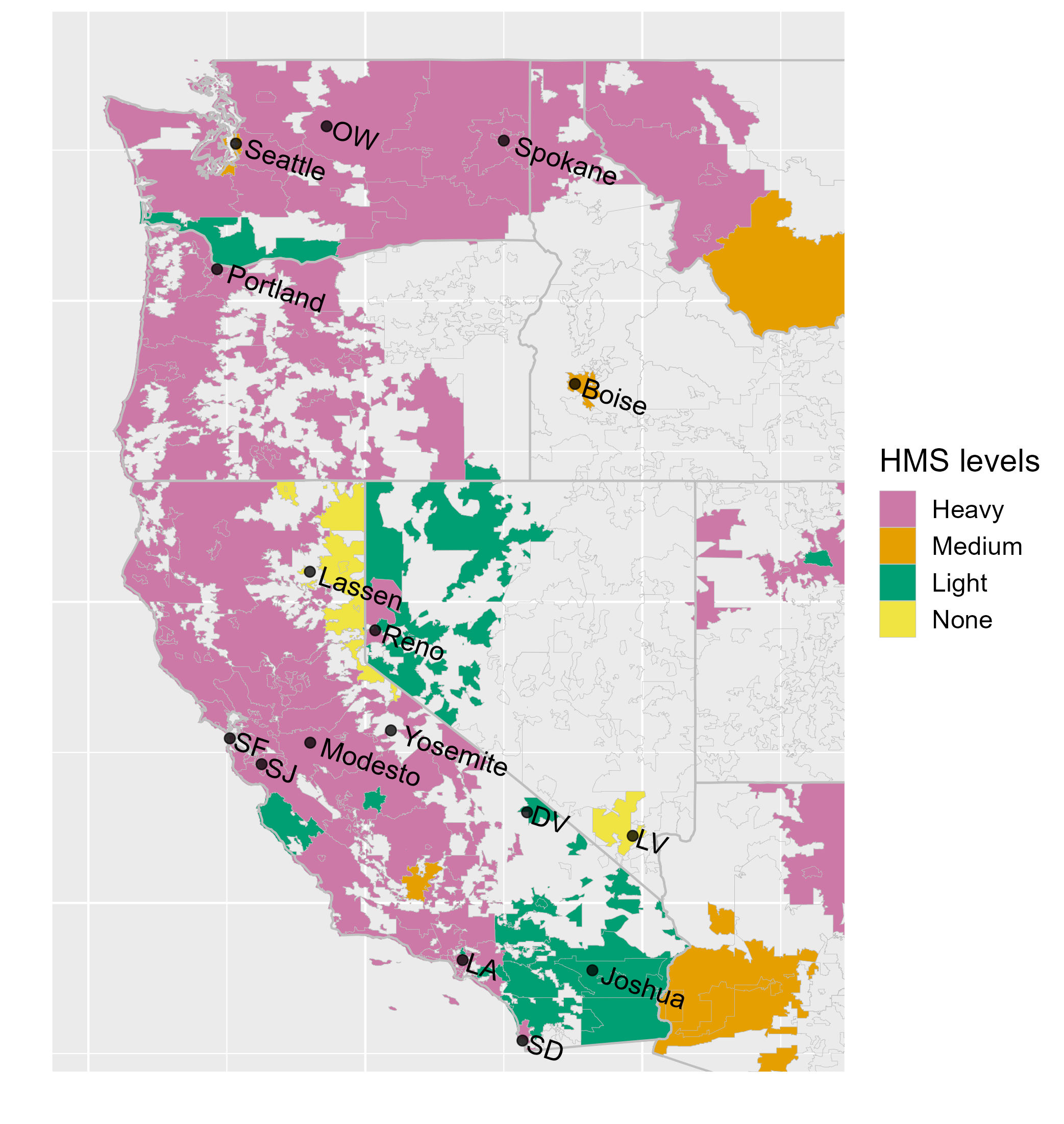} 
\includegraphics[width=0.4\textwidth]{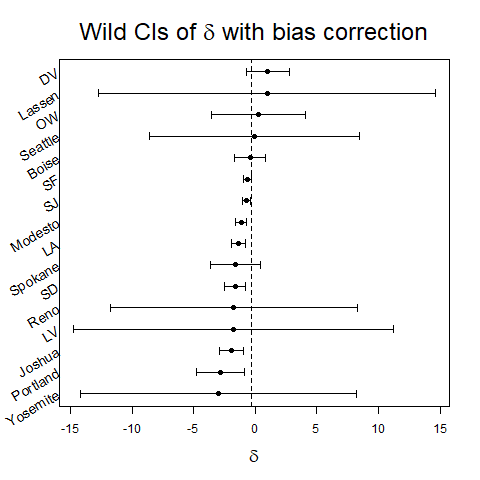} 
\par\end{center}
\caption{Location (left) and 95\% confidence intervals of treatment effects
(right) for the selected cities and national parks. The vertical black
dash line is the global treatment effect estimate. \label{fig:CIs-for-the-cities}}
\end{figure}

We fit the local model to several cities including Spokane, Portland,
Seattle, San Diego (SD), San Francisco (SF), Los Angeles (LA), Reno
and Las Vegas (LV), as well as prominent national parks including
Okanogan-Wenatchee National Forest (OW), Boise, Yosemite National
Park (Yosemite), Death Valley National Park (DV), Joshua Tree National
Park (Joshua), Lassen National Forest (Lassen). Figure \ref{fig:CIs-for-the-cities}
shows a forest plot of the point estimates and confidence intervals
for $\delta$. The results align with our discoveries in Section \ref{subsec:Local-constant-estimation} that a majority of users in southwest locations (e.g., LA, SD, and Joshua) have significant beneficial treatment effects, whereas such pronounced benefits are uncommon in the northwest. The right panel of Figure \ref{fig:CIs-for-the-cities} provides further evidence with location-specific confidence intervals, confirming the existence of a spatially diverse treatment effect.

\section{Discussion \label{sec:Discussion}}

It has become increasingly feasible to evaluate treatment and intervention
strategies on public health as more health and activity data are collected
via mobile phones, wearable devices, and smartphone applications.
We establish a new framework of spatially and time-varying causal
effect models. This provides a theoretical foundation to utilize emerging
smartphone application data to draw causal inference of interventions
on health outcomes. Our approach does not require specifying the full
distribution of the covariate, treatment, and outcome processes. Moreover,
our method achieves a double robustness property requiring the correct
specification of either the model for the outcome mean or the model
for the treatment process. The key underpinning assumption is sequential
treatment randomization, which holds if all variables are measured
that are related to both treatment and outcome. Although essential,
it is not verifiable based on the observed data but relies on subject
matter experts to assess its plausibility.

The goal of the Smoke Sense citizen science study is to engage the
participants on the issue of wildfire smoke as a health risk and facilitate
the adaptation of health-protective measures. Our new analysis framework
reveals that there is a spatially varying health benefit and that
the global model underestimates the treatment effect in areas with
the highest exposure to wildland fire smoke. This new knowledge
obtained from the spatial analysis may also help Smoke Sense scientists
and developers improve the app by targeting different people with
different messaging.

There are several directions for future work.
First, the study's conclusion is restricted to Smoke Sense participants and may not apply to the general population. Generalizing these findings to a broader population is an interesting topic, as discussed by \cite{lee2022doubly},
\cite{lee2023improving},
\cite{lee2024genrct}, and
\cite{lee2024transporting}.
Second, this work focuses on structural nested ``mean'' models for continuous or approximately
continuous outcomes. It is important to continue the development of
local causal models to accommodate different types of outcomes. For
example, we can consider the structural nested failure time models
for a time-to-event outcome \citep{yang2018semiparametric}. Third,
the current framework relies on the sequential treatment randomization
assumption. \citet{yang2017sensitivity} relaxed this assumption by defining a bias function that
quantifies the impact of unmeasured confounding and developed a modified
estimator for the class of global SNMMs. Additional work is necessary
to assess the impact of possible uncontrolled confounding for the
new class of local SNMMs. Alternatively, \cite{guan2023spectral} introduced a spectral perspective that presents a novel approach to addressing unmeasured spatial confounders, which could be explored in our context.
Finally, with the app is continuously collecting
more data from the users, one of the interesting future directions
would be incorporating the covariates into the treatment effect model
and estimate the optimal personalized behavior recommendations.

\section*{Acknowledgment}
	We thank the Editor, an Associate Editor, and two Reviewers for their insightful comments and helpful suggestions on the earlier version of this paper. We would also like to thank Linda Wei for the help with the data.  

\section*{Supplementary Material}
Supplementary materials available online include all technical proofs, additional simulation results, other implementation details, and R functions. 

\section*{Data Availability}
Data subject to third party restrictions.

\section*{Funding}
This research is supported by NSF DMS 1811245, NCI P01 CA142538, NIA 1R01AG066883, and NIEHS 1R01ES031651. 

\section*{Conflict of Interests}
None declared.


\bibliographystyle{chicago}
\bibliography{refs}

\listoffigures
\end{document}


The supplementary material includes technical details, proofs, and
additional details for the real data application.

\section{Time-varying confounding and structural nested mean models}\label{sec:SNMM_additional}

In the presence of time-varying confounders, unique challenges arise in longitudinal studies when drawing valid causal inferences, as the associations between treatment and outcome are distorted over time \citep{yang2016goodness, yang2018sensitivity, yang2022semiparametric}. Consider the \textit{directed acyclic graph} (DAG) depicted in Figure \ref{fig:DAG_eg} to illustrate the causal relations between these variables for simplicity. DAGs are the most common tools used to represent causality. For example, the directed edge from $A_1$ to $Y_1$ indicates that $A_1$ is a direct cause of $Y_1$. Suppose we aim to examine the causal effects of $A_1$ and $A_2$ on $Y_2$ at time $t_2$. First, we observe that $X_2$ acts as the confounder of $A_2$ and $Y_2$ due to the direct causes from $X_2$ to both $A_2$ and $Y_2$. Consequently, controlling for $X_2$ is necessary to study the causal effect of $A_2$ on $Y_2$, which is formulated by $\E\{Y_2^{(a_1, a_2)} - Y_2^{(a_1, 0)}\mid A_1=a_1, X_1, X_2\}$. However, since $X_2$ is an intermediate variable of $A_{1}$ and $Y_2$, lying in the causal pathway between $A_1$ and $Y_2$, it should not be controlled. This is because $X_2$ may partly explain away the causal effect of $A_1$ on $Y_2$. Thus, the causal effect of $A_1$ on $Y_2$ should be formulated as $\E\{Y_2^{(a_1)} - Y_2^{(0)}\mid X_1\}$ without controlling $X_2$. However, using standard regression methods to investigate time-varying treatment effects may exhibit biases after adjusting for both $X_1$ and $X_2$ \citep{robins2009estimation}.
\begin{figure}[htbp]
\centering
\begin{tikzpicture}
    \node[state] (A1) at (0,0) {$A_{1}$};
    \node[state] (Y1) [right =of A1] {$Y_{1}$}; 
    \node[state] (X1) [above right = of A1]{$X_1$};
\node[state] (A2) [right =of Y1] {$A_{2}$};
    \node[state] (X2) [above right =of A2] {$X_{2}$};
\node[state] (Y2) [right =of A2] {$Y_{2}$};
    \path (A1) edge (Y1);
    \path (X2) edge (A2);
    \path (A2) edge (Y2);
    \path (X1) edge (A1);
    \path (X1) edge (Y1);
    \path (Y1) edge (A2);
    \path (X2) edge (Y2);
    \path (A1) edge (X2);
\end{tikzpicture}
\caption{A directed acyclic graph illustrating the relationships among $\overline{A}$, $\overline{X}$ and $\overline{Y}$. \label{fig:DAG_eg}}
\end{figure}

\section{The optimal choice of $q_{k,m}(\overline{L}_{m})$ in (\ref{revision-eq:opt ef})}

\citet{robins1994correcting} established the semiparametric efficiency
theory of the structural nested mean models in the non-spatial setting.
Following \citet{robins1994correcting}, we consider 
\begin{multline}
\left(\begin{array}{c}
q_{m,m}(\overline{L}_{m})\\
\vdots\\
q_{K,m}(\overline{L}_{m})
\end{array}\right)=\left(\E\left[\frac{\partial}{\partial\psi(\bs)}\left.\left(\begin{array}{c}
H_{m}^{(\overline{A}_{m-1})}\{\psi^{*}(\bs)\}\\
\vdots\\
H_{K}^{(\overline{A}_{m-1})}\{\psi^{*}(\bs)\}
\end{array}\right)\right\vert A_{m}=1,\overline{L}_{m}\right]\right.\\
-\left.\E\left[\frac{\partial}{\partial\psi(\bs)}\left.\left(\begin{array}{c}
H_{m}^{(\overline{A}_{m-1})}\{\psi^{*}(\bs)\}\\
\vdots\\
H_{K}^{(\overline{A}_{m-1})}\{\psi^{*}(\bs)\}
\end{array}\right)\right\vert A_{m}=0,\overline{L}_{m}\right]\right)\var\left[\left.\left(\begin{array}{c}
H_{m}^{(\overline{A}_{m-1})}\{\psi^{*}(\bs)\}\\
\vdots\\
H_{K}^{(\overline{A}_{m-1})}\{\psi^{*}(\bs)\}
\end{array}\right)\right\vert \overline{L}_{m}\right]^{-1},\label{eq:opt-q}
\end{multline}
for $m=1,\ldots,K$. Assuming further that $\E\left[H_{k}^{(\overline{A}_{m-1})}\{\psi^{*}(\bs)\}\mid A_{m},\overline{L}_{m}\right]=\E\left[H_{k}^{(\overline{A}_{m-1})}\{\psi^{*}(\bs)\}\mid\overline{L}_{m}\right]$
for all $m\leq k$, this choice of $q_{k,m}(\overline{L}_{m})$ leads to
the solution to $\mathbb{P}_{n}\left[G\{V;\psi(\bs)\}\right]=0$ having
the smallest asymptotic variance compared to all other choices.

\section{Proof of Theorem \ref{revision-thm:asymp}}

To simplify exposition, we introduce additional notation. We use $V_{1}\cong V_{2}$
to denote the two random variables $V_{1}$ and $V_{2}$ are asymptotically
equivalent up to a smaller order term. Let $M_{1}$ and $M_{2}$ be
$m_{1}\times l_{1}$ and $m_{2}\times l_{2}$ matrices, respectively,
then $M_{1}\otimes M_{2}$ is the kronecker product of $M_{1}$ and
$M_{2}$, defined as a $(m_{1}m_{2})\times(l_{1}l_{2})$ matrix by
multiplying each component of $M_{1}$ by $M_{2};$ e.g., if $M_{1}$
is a $3\times1$ matrix $d(\bs^{*}-\bs)$, 
\[
M_{1}\otimes M_{2}=\left(\begin{array}{c}
1\\
s_{1}^{*}-s_{1}\\
s_{2}^{*}-s_{2}
\end{array}\right)\otimes M_{2}=\left(\begin{array}{c}
M_{2}\\
(s_{1}^{*}-s_{1})M_{2}\\
(s_{2}^{*}-s_{2})M_{2}
\end{array}\right).
\]
Let $c_{K}(r)=\int\bs^{r}K(\bs)\de\bs$, with $c_{K}(0)=c_{K}(2)=1$,
and $\gamma_{K}(r)=\int\bs^{r}K^{2}(\bs)\de\bs$. For a function of
$\bs$, $h(\bs)$, denote the second order partial derivative as 
\[
D_{h(\bs)}^{(2)}(\bs)=\left(\begin{array}{cc}
\frac{\partial^{2}h(\bs)}{\partial s_{1}^{2}} & \frac{\partial^{2}h(\bs)}{\partial s_{1}\partial s_{2}}\\
\frac{\partial^{2}h(\bs)}{\partial s_{2}\partial s_{1}} & \frac{\partial^{2}h(\bs)}{\partial s_{2}^{2}}
\end{array}\right).
\]

To reflect the dependence of the estimating function on estimators
of nuisance functions, denote

\[
\widehat{G}\{V;\psi(\bs)\}=\sum_{m=1}^{K}\sum_{k=m}^{K}q_{k,m}(\overline{L}_{m})\left[H_{k}^{(\overline{A}_{m-1})}\{\psi(\bs)\}-\widehat{\mu}_{m,k}(\overline{L}_{m})\right]\{A_{m}-\widehat{e}(\overline{L}_{m})\}.
\]
Then, $\widehat{\psi}_{\tau}(\bs^{*})$ is the solution to 
\begin{equation}
\mP_{n}\left[\omega_{\tau}(||\bs^{*}-\bs||)d(\bs^{*}-\bs)\otimes\widehat{G}\{V;\widetilde{\psi}_{\lp}(\bs^{*};\phi)\}\right]=0.\label{eq:GWE-1}
\end{equation}

Let $\widetilde{\mu}_{m,k}(\overline{L}_{m})$ and $\widetilde{e}(\overline{L}_{m})$
be the probability limits of $\widehat{\mu}_{m,k}(\overline{L}_{m})$ and
$\widehat{e}(\overline{L}_{m})$, respectively. If the working model for
$\mu_{m,k}(\overline{L}_{m})$ is correctly specified, $\widetilde{\mu}_{m,k}(\overline{L}_{m})=\mu_{m,k}(\overline{L}_{m})$.
If the working model for $e(\overline{L}_{m})$ is correctly specified,
$\widetilde{e}(\overline{L}_{m})=e(\overline{L}_{m})$. Define 
\[
\widetilde{G}\{V;\psi(\bs)\}=\sum_{m=1}^{K}\sum_{k=m}^{K}q_{k,m}(\overline{L}_{m})\left\{ H_{k}^{(\overline{A}_{m-1})}\{\psi(\bs)\}-\widetilde{\mu}_{m,k}(\overline{L}_{m})\right\} \{A_{m}-\widetilde{e}(\overline{L}_{m})\}.
\]
Let $\widetilde{\chi}\{V;\psi(\bs)\}=\partial\widetilde{G}\{V;\psi(\bs)\}/\partial\psi(\bs).$
Also, define $B(\bs^{*})=\E[\widetilde{\chi}\{V;\psi(\bs^{*})\}],$
and $C(\bs^{*})=\E\left[\widetilde{G}\{V;\psi(\bs^{*})\}\widetilde{G}\{V;\psi(\bs^{*})\}^{\T}\right].$

To study the asymptotic bias and variance of $\widehat{\phi}_{\tau,0}(\bs^{*})$,
following \citet{carroll1998local}, we use a Taylor expansion to
obtain 
\begin{equation}
\widehat{\phi}_{\tau,0}(\bs^{*})-\psi^{*}(\bs^{*})\cong-e_{1}^{\T}\left\{ B_{n,\tau}(\bs^{*})\right\} ^{-1}\widehat{D}_{n,\tau}(\bs^{*}),\label{eq:car1}
\end{equation}
where 
\begin{eqnarray*}
B_{n,\tau}(\bs^{*}) & = & \mathbb{P}_{n}\omega_{\tau}(||\bs^{*}-\bs||)d(\bs^{*}-\bs)\otimes\widetilde{\chi}\{V;\psi(\bs)\},\\
\widehat{D}_{n,\tau}(\bs^{*}) & = & \mathbb{P}_{n}\omega_{\tau}(||\bs^{*}-\bs||)d(\bs^{*}-\bs)\otimes\widehat{G}\{V;\psi(\bs)\}.
\end{eqnarray*}

Because $\widehat{\E}[H_{k}^{(\overline{A}_{m-1})}\{\psi(\bs)\}\mid\overline{L}_{m}]=\widetilde{\E}[H_{k}^{(\overline{A}_{m-1})}\{\psi(\bs)\}\mid\overline{L}_{m}]+o_{P}(n^{-1/2})$
and $\widehat{e}(\overline{L}_{m})=\widetilde{e}(\overline{L}_{m})+o_{P}(n^{-1/2})$,
which coverage faster then kernel weighting estimators, we can further
express (\ref{eq:car1}) as

\begin{eqnarray*}
\widehat{\phi}_{\tau,0}(\bs^{*})-\psi^{*}(\bs^{*}) & \cong & -e_{1}^{\T}\left\{ B_{n,\tau}(\bs^{*})\right\} ^{-1}\widetilde{D}_{n,\tau}(\bs^{*}),
\end{eqnarray*}
where 
\begin{eqnarray*}
\widetilde{D}_{n,\tau}(\bs^{*}) & = & \mathbb{P}_{n}\omega_{\tau}(||\bs^{*}-\bs||)d(\bs^{*}-\bs)\otimes\widetilde{G}\{V;\psi(\bs)\}.
\end{eqnarray*}

To study the asymptotic bias and variance of $\widehat{\phi}_{\tau,0}(\bs^{*})-\psi(\bs^{*})$,
we study the two terms $B_{n,\tau}(\bs^{*})$ and $\widetilde{D}_{n,\tau}(\bs^{*})$
separately. First, note that the spatial location $\bs$ is a two-dimensional
variable, following \citet{ruppert1994multivariate}, we obtain

\[
B_{n,\tau}(\bs^{*})\cong f(\bs^{*})\left(\begin{array}{c}
c_{K}(2)\\
0\\
0
\end{array}\right)\otimes B(\bs^{*}).
\]
Then, it suffices to focus on the first $p\times p$ block of $\widetilde{D}_{n,\tau}(\bs^{*})$.

To evaluate the asymptotic bias of the first $p\times p$ block of
$\widetilde{D}_{n,\tau}(\bs^{*})$, it is important to note that if
either model for $\mu_{m,k}(\overline{L}_{m})$ or $e(\overline{L}_{m})$ is
correctly specified, $\E[\widetilde{G}\{V;\psi(\bs^{*})\}]=0$. Then,
we have 
\begin{eqnarray*}
 &  & \E\omega_{\tau}(||\bs^{*}-\bs||)\widetilde{G}\{V;\psi_{\lp}(\bs;\phi)\}\\
 & = & \E\omega_{\tau}(||\bs^{*}-\bs||)\left[\widetilde{G}\{V;\psi_{\lp}(\bs;\phi)\}-\widetilde{G}\{V;\psi(\bs^{*})\}\right]\\
 & = & -\E\omega_{\tau}(||\bs^{*}-\bs||)\widetilde{\chi}\{V;\psi(\bs^{*})\}\frac{1}{2}(\bs-\bs^{*})^{\T}D_{\psi}^{(2)}(\bs^{*})(\bs-\bs^{*})+o_{P}(\tau^{2})\\
 & = & -\frac{1}{2}c_{K}(2)B(\bs^{*})\trace\left\{ D_{\psi}^{(2)}(\bs^{*})\right\} \tau^{2}+o_{P}(\tau^{2}),
\end{eqnarray*}
where the first equality follows because $\E[\omega_{\tau}(||\bs^{*}-\bs||)\widetilde{G}\{V;\psi(\bs^{*})\}]=0$
if either nuisance model is correctly specified.

To evaluate the asymptotic variance of the first $p\times p$ block
of $\widetilde{D}_{n,\tau}(\bs^{*})$, we have 
\begin{eqnarray*}
 &  & \mathbb{P}_{n}\left(\omega_{\tau}(||\bs^{*}-\bs||)^{2}\left[\widetilde{G}\{V;\psi_{\lp}(\bs;\phi)\}\widetilde{G}\{V;\psi_{\lp}(\bs;\phi)\}^{\T}\right]\right)\\
 & = & n\tau^{2}f_{\bs}(\bs^{*})\gamma_{K}(0)C(\bs^{*})+o_{p}(\tau^{2}).
\end{eqnarray*}
Combining all results, we show the result in Theorem \ref{revision-thm:asymp}
with 
\begin{eqnarray*}
\mathcal{G}_{b}\{\bs^{*},K,\psi(\bs^{*})\} & = & \frac{1}{2}c_{K}(2)\trace\left\{ D_{\psi}^{(2)}(\bs^{*})\right\} ,\\
\mathcal{G}_{v}\{\bs^{*},K,\psi(\bs^{*})\} & = & \gamma_{K}(0)\left\{ B(\bs^{*})\right\} ^{-1}C(\bs^{*})\left\{ B(\bs^{*})\right\} ^{-1,\T}.
\end{eqnarray*}

\section{Technical details with informative non-responses}

In the geographically weighted local polynomial framework, an additional
complication arises due to nuisance function estimation of $\mu_{m,k}(\overline{L}_{m})$
and $e(\overline{L}_{m})$ in the presence of non-responses, which now
requires weighting. In this setting, the proposed estimation procedure
proceeds as follows. 
\begin{description}
\item [{Step$\ $S.0.}] Fit a non-response model by solving (\ref{revision-eq:GMM}),
denoted by $\pi_{m}(V_{m};\widehat{\eta})$ and $\widehat{\pi}_{m}$
for shorthand. 
\item [{Step$\ $S.1.}] Fit a propensity score model weighted by the inverse
of the product of non-response probabilities over the required time
points for observations. For example, if $e(\overline{L}_{m})$ depends
only on $X_{m}$, then fitting the model $e(X_{m})$ based on $\{(A_{m},X_{m}):0\leq m\leq K,R_{m}=1\}$
with weight $R_{m}/\widehat{\pi}_{m}$. If $e(\overline{L}_{m})$ depends
only on $(A_{m-1},X_{m})$, then fitting the model $e(A_{m-1},X_{m})$
based on $\{(A_{m-1},A_{m},X_{m}):0\leq m\leq K,R_{m-1}R_{m}=1\}$
with weight $R_{m-1}R_{m}/(\widehat{\pi}_{m-1}\widehat{\pi}_{m})$.
Denote the estimated propensity score by $\widehat{e}(\overline{L}_{m})$. 
\item [{Step$\ $S.2.}] For the treatment effect $\psi^{*}(\bs^{*})$ at
a given location $\bs^{*}$, where $\bs^{*}$ varies over all observed
locations, obtain a preliminary estimator $\widehat{\phi}_{\tau}^{[0]}(\bs^{*})$
by solving 
\[
\mP_{n}\omega_{\tau}(||\bs^{*}-\bs||)d(\bs^{*}-\bs)\otimes G_{\ipw}^{[0]}\{V;\widetilde{\psi}_{\lp}(\bs;\phi)\}=0,
\]
for $\phi,$ where 
\[
G_{\ipw}^{[0]}\{V;\psi(\bs)\}=\sum_{m=1}^{K}\sum_{k=m}^{K}\widehat{\omega}_{m:k}(\overline{R})q_{k,m}^{[0]}(\overline{X}_{m})H_{k}^{(\overline{A}_{m-1})}\{\psi(\bs)\}\{A_{m}-e(\overline{L}_{m})\},
\]
and $\widehat{\omega}_{m:k}(\overline{R})=\prod_{l=m}^{k}\{R_{l}\widehat{\pi}_{l}(V)^{-1}\}.$
Then, $\widehat{\phi}_{\tau,0}^{[0]}(\bs^{*})$, the first $p$-vector
in $\widehat{\phi}_{\tau}^{[0]}(\bs^{*})$, is a preliminary estimator
of $\psi(\bs^{*})$. With $\widehat{\phi}_{\tau,0}^{[0]}(\bs^{*})$,
we approximate $H_{k}^{(\overline{A}_{m-1})}\{\psi(\bs^{*})\}$ by$\widehat{H}_{k}^{(\overline{A}_{m-1})}\{\psi(\bs^{*})\}=H_{k}^{(\overline{A}_{m-1})}\left\{ \widehat{\phi}_{\tau,0}^{[0]}(\bs^{*})\right\} $.
Using the pseudo outcome $\widehat{H}_{k}^{(\overline{A}_{m-1})}\{\psi(\bs^{*})\}$
and $\overline{L}_{m}$, fit an outcome mean model using the weighting
strategy in Step S.1, denoted by $\widehat{\mu}_{m,k}(\overline{L}_{m})$. 
\item [{Step$\ $S.3.}] Obtain $\widehat{\phi}_{\tau}(\bs^{*})$ by solving
(\ref{revision-eq:GWE}) with $G\{V;\widetilde{\psi}_{\lp}(\bs^{*};\phi)\}$
replaced by $G_{\mathrm{ipw}}\{V;\widetilde{\psi}_{\lp}(\bs^{*};\phi)\}$,
$\pi_{m}(V_{m})$, $e(\overline{L}_{m})$ and $\mu_{m,k}(\overline{L}_{m})$
replaced by their estimates $\widehat{\pi}_{m}$, $\widehat{e}(\overline{L}_{m})$
and $\widehat{\mu}_{m,k}(\overline{L}_{m})$. 
\end{description}

\section{Additional simulation results\label{sec:sim_additional}}
In this section, we will provide additional results from the simulation studies. Table \ref{tab:MISEs} evaluates the performance of the estimated function $\widehat{\psi}(\boldsymbol{s})$ by presenting the mean integrated squared errors (MISE), defined by $100^{-1}\sum_{j}\{\psi^*(\boldsymbol{s}_j)-\widehat{\psi}(\boldsymbol{s}_j)\}^2$, over a $10\times 10$ grid of spatial locations when the variance of $\epsilon_k$ is $0.25$ and $3$. The $10\times 10$ grid is generated by uniform sampling over a specific range of $s_1$ and $s_2$. For scenarios (S1) and (S2), $s_1$ and $s_2$ take values from $0$ to $1$, whereas $s_1$ and $s_2$ range from $37.5$ to $45.0$ and $-121.9$ to $-120.7$ in scenarios (S3) and (S4) to mimic the location features of the real data, respectively. Scenario (S4) with $\psi^*(\bs)=(s_1+s_2)/100$ is added to benchmark the performance when we mirror the features of the Smoke Sense data. The small MISEs over all scenarios regardless of the magnitude of the error term indicate good function performances over the target grid.

\begin{table}
\caption{\label{tab:MISEs} Mean integrated squared errors (MISE) for scenarios (S1)-(S4) over a $10\times 10$ grid when $\mathrm{var}(\epsilon_k)=0.25$ and $0.3$.}
\vspace{0.25cm}
    \begin{tabular}{cccccc}
    \toprule
         &$\rm{var}(\epsilon_k)$ & (S1) & (S2) & (S3) & (S4) \\
    \midrule
         MISE $(\times 10^{-3})$ & 0.25 & 1.48 & 3.33 & 2.00 & 0.77\\
         & 3 & 1.50 & 4.93 & 8.86 &5.63\\
    \bottomrule
    \end{tabular}
\end{table}

Figure \ref{fig:mc-tau-avg} displays the Monte Carlo average of the bandwidth $\tau$ selected by 5-fold cross-validation over a $10\times 10$ grid for all four scenarios when $\rm{var}(\epsilon_k)=0.25$. Intuitively, $\tau$ gauges the flatness of spatial treatment effect $\psi^*(\boldsymbol{s})$; a large value $\tau$ may indicate a flat surface of $\psi^*(\boldsymbol{s})$ locally as more neighbors are incorporated in the geographically weighted estimation. For example in (S1) where the treatment effect function $\psi^*(\boldsymbol{s})$ is $\exp(s_1+s_2)$, the 5-fold CV tends to select large values of $\tau$ for the left-bottom grid but small values for the right-top grid. As the treatment effects are rather flat in the left-bottom locations for (S1) (i.e., small $\Delta \psi^*(\boldsymbol{s})/\Delta \boldsymbol{s}$), it is reasonable to choose a large bandwidth window $\tau$ to leverage more information from their neighborhood; when the treatment effects become steeper for the right-top locations (i.e., large $\Delta \psi^*(\boldsymbol{s})/\Delta \boldsymbol{s}$), smaller bandwidth values $\tau$ are more desired. 

\begin{figure}
    \centering
\includegraphics[width=.75\linewidth]{figures/tau_S1_to_S4_sd0_25.pdf}
    \caption{Monte Carlo average of $\tau_{\text{cv}}$ selected by 5-fold cross-validation over $10\times 10$ grid when $\rm{var}(\epsilon_k)=0.25$. \label{fig:mc-tau-avg}}
\end{figure}

\section{Details for real data application}

In this section, we first provide the scientific meanings of each selected variable and give a preliminary exploratory data analysis for the Smoke Sense data. Next, we provide several sensitivity analyses including changing the nuisance models, the imputation method, and the functional form of the treatment effect model. Furthermore, we present the complete estimation results for $\psi^{*}(\boldsymbol{s})$ with the confidence intervals at the 28 locations and report the selected bandwidth $\tau_{\text{cv}}$ at each location. Lastly, we provide a model diagnostic statistic to measure how well the proposed treatment effect model fits the data.

\subsection{Exploratory data analysis\label{sec:EDA}}

First, we describe the meanings of covariates, outcome, and treatment of the Smoke Sense data in detail. Starting with the definition of the treatment, we consider the binary variable: $A_{k}=1$ if the user took strong protective behaviors: stayed indoor and used a respirator mask N95 or similar with extra filtration at the same time, or stayed indoors and used an air cleaner (room HEPA air purifier or similar) at the same time; and $A_{k}=0$ if the subject
did not take any behavior or took any other mild protective behaviors
such as reducing normal outdoor recreation, keeping windows and doors
closed, etc.

For the outcome, we consider the number of symptoms as follows: 
\begin{enumerate}
\item Anxiety 
\item Asthma attack 
\item Chest pain 
\item Coughing trouble breathing normally shortness of breath 
\item Ear or other viral infections 
\item Fast or irregular heart beat 
\item High blood pressure 
\item Irritated sinuses congestion 
\item Runny or stuffy nose 
\item Scratchy throat 
\item Stinging itching or watery eyes 
\item Tiredness dizziness or similar 
\item Trouble sleeping 
\item Wheezing 
\item Other 
\end{enumerate}
Figure \ref{fig:Y_EDA} presents the geographic plot and the histogram of the number of user-reported adverse health symptoms, excluding zero values, from September 2018 to the end of 2019 for the purpose of exploratory data analysis. The histogram suggests that it is reasonable to treat the observed outcomes as continuous variables. 
\begin{figure}
        \centering
        \includegraphics[width = .9\linewidth]{figures/EDA_Y_all.pdf}
        \caption{\label{fig:Y_EDA} The geographic plot and the histogram overlaid with the kernel density of the observed number of user-reported adverse health symptoms.}
    \end{figure}

We consider the following baseline variables with summary statistics
reported in Tables \ref{tab:summary baseline1} - \ref{tab:summary baseline3}: 
\begin{enumerate}
\item Gender $\in\{\text{Female, Not female}\}$. 
\item Race $\in\{\text{White, Not white}\}$. 
\item Age $\in\{18-29,30-39,40-49,50-64,65+\}$. 
\item Highest education level $\in\{$(1) High school degree, GED, or less;
(2) Technical school, trade or vocational training, or associate degree;
(3) Bachelors, masters, doctorate, or professional degree$\}$. 
\item Health status: Would you say your own health in general is: \{Poor,
Fair, Good, Very Good, Excellent\}. 
\item No considering to reduce smoke: \{1: would not consider reducing exposure;
0: otherwise\}. 
\item No smoke-reducing tool: \{1: no reducing smoke tools; 0: do have reducing
smoke tools\}. 
\item Activeness: On average, when you are outside, how active are you:
\{Not Very Active, Mild (walking, standing), Moderate (regular jog,
gardening), Very Active (run, bike daily, work outdoors)\}. 
\item Common smoke: Smoke from wildfires is a common occurrence where I
live: \{Strongly disagree, Somewhat disagree, Neither agree nor disagree,
Somewhat agree, Strongly agree\}. 
\item Smoke impact: A few hours of wildfire smoke in the air can impact
my health: \{Strongly disagree, Somewhat disagree, Neither agree nor
disagree, Somewhat agree, Strongly agree\}. 
\item Possible to reduce smoke: It is possible for me to reduce my wildfire
smoke exposure: \{Strongly disagree, Somewhat disagree, Neither agree
nor disagree, Somewhat agree, Strongly agree\}. 
\item Alert help: Information alerts are likely to help me reduce my exposure
to wildfire smoke: \{Strongly disagree, Somewhat disagree, Neither
agree nor disagree, Somewhat agree, Strongly agree\}. 
\item Symptom 1: the number of symptoms in the baseline characteristics
$\in\{0,1,\dots,5\}$; the symptoms include: \{Coughing trouble breathing
wheezing asthma attacks or similar, High blood pressure chest pain
or tightness rapid or irregular heartbeat or similar, Runny or stuffy
nose irritated sinuses or similar, Stinging eyes scratchy throat or
similar, Tiredness headaches or similar\}. 
\item Symptom 2: the number of symptoms in the baseline characteristics
$\in\{0,1,\dots,8\}$; the symptoms include; \{Allergies related to
the upper respiratory tract eyes and ears, Asthma, Chronic Obstructive
Pulmonary Disease, Hypertension or high blood pressure, Other chronic
disease, Other heart disease, Other respiratory disease, Type II diabetes
metabolic syndrome or obesity\}. 
\end{enumerate}
We consider the following time-varying variables: 
\begin{enumerate}
\item Did you take these actions before you first had symptoms: \{Does not
apply, No, Yes\}. 
\item Air quality yesterday: How was your air quality yesterday: \{Dangerous,
Very poor, Somewhat poor, Moderate, Good\}. 
\item Recent smoke experience: Have you experienced smoke recently: \{No,
I didn't experience smoke; Yes, I experienced smoke in the last month
but not this week; Yes, I experienced smoke this week\}. 
\item Feeling status: On the day you felt the worst over the past week,
how much discomfort did you feel compared to your normal state: \{Very
severe discomfort, Severe discomfort, Moderate discomfort, Mild discomfort,
Same as normal\}. 
\item Smoke perspective: To what extent do you agree with this statement
thinking about the past week: I was sensitive to wildfire smoke: \{Strongly
disagree, Somewhat disagree, Neither agree nor disagree, Somewhat
agree, Strongly agree\}. 
\item Action perspective: To what extent do you agree with this statement:
If there is a serious smoke event next week, it will be worth it to
take actions to reduce my exposure: \{Strongly disagree, Somewhat
disagree, Neither agree nor disagree, Somewhat agree, Strongly agree\}. 
\item Days of feeling smoke inside: Did you smell smoke inside of your
home: \{Not at all, 1-2 days, 3+ days\}. 
\item Days of feeling smoke outside: Did you smell smoke outside your home,
workplace, school during, this, time: \{Not at all, 1-2 days, 3+ days\}. 
\item Days visibility impacted: How many days was your visibility impacted
by smoke: \{Not at all, 1-2 days, 3+ days\}. 
\item Comparison with the worst day: On a cloudy day visibility ranges from
7 to 10 miles; On the worst day over the past week, how did your visibility
compare: the larger the more visibility: \{Almost no visibility, Extremely
worse, Much worse, A little worse, About the same\}. 
\item HMS: Weekly maximum smoke density recorded $\in$\{0, 5, 16, and 27\}
$(\mu g/m^{3})$. 
\item The number of actions taken in the past. 
\end{enumerate}
\begin{table}
\caption{\label{tab:summary baseline1}Summary statistics of baseline variables (part I)}
\vspace{0.25cm}
 \resizebox{0.9\textwidth}{!}{ %
\begin{tabular}{lcc}
\toprule 
$\mathbf{n = 1882}$  & \textbf{{Number } }  & \textbf{{Percent} }\tabularnewline
 & \textbf{of Responses}  & \textbf{of Responses (\%)}\tabularnewline
\midrule 
\multicolumn{3}{l}{\textbf{{Gender}}}\tabularnewline
\midrule 
Female  & 961  & 51.06 \tabularnewline
Male  & 615  & 32.68 \tabularnewline
N/A  & 306  & 16.26 \tabularnewline
\midrule 
\multicolumn{3}{l}{\textbf{{Race}}}\tabularnewline
\midrule 
White  & 1253  & 66.58 \tabularnewline
Asian/Pacific Islander  & 79  & 4.20 \tabularnewline
Hispanic/Latino  & 71  & 3.77 \tabularnewline
Native American  & 18  & 0.96 \tabularnewline
African-American/Black  & 19  & 1.01 \tabularnewline
Other  & 107  & 5.69 \tabularnewline
N/A  & 335  & 17.80 \tabularnewline
\midrule 
\multicolumn{3}{l}{\textbf{{No smoke-reducing tool}}}\tabularnewline
\midrule 
0: do have reducing smoke tools  & 1437  & 76.35 \tabularnewline
1: no reducing smoke tools 0:  & 82  & 4.36 \tabularnewline
N/A  & 363  & 19.29 \tabularnewline
\midrule 
\multicolumn{3}{l}{\textbf{{No considering to reduce smoke}}}\tabularnewline
\midrule 
0: otherwise  & 1410  & 74.92 \tabularnewline
1: would not consider reducing exposure  & 21  & 1.11 \tabularnewline
N/A  & 451  & 23.96 \tabularnewline
\midrule 
\multicolumn{3}{l}{\textbf{{Age}}}\tabularnewline
\midrule 
$18-29$  & 229  & 12.18 \tabularnewline
$30-39$  & 343  & 18.23 \tabularnewline
$40-49$  & 324  & 17.22 \tabularnewline
$50-64$,  & 416  & 22.10 \tabularnewline
65+  & 256  & 13.60 \tabularnewline
N/A  & 314  & 16.68 \tabularnewline
\midrule 
\multicolumn{3}{l}{\textbf{{Activeness: On average, when you are outside, how active
are you}}}\tabularnewline
\midrule 
Not Very Active  & 702  & 37.30 \tabularnewline
Mild (walking, standing)  & 519  & 27.58 \tabularnewline
Moderate (regular jog, gardening)  & 59  & 3.13 \tabularnewline
Very Active(run, bike daily, work outdoors)  & 244  & 12.96 \tabularnewline
N/A  & 358  & 19.02 \tabularnewline
\midrule 
\multicolumn{3}{l}{\textbf{{Highest education level}}}\tabularnewline
\midrule 
High school degree, GED, or less  & 182  & 9.67 \tabularnewline
Technical school, trade or vocational training, or associate degree  & 253  & 13.44 \tabularnewline
Bachelors, masters,doctorate, or professional degree  & 1097  & 58.29 \tabularnewline
N/A  & 350  & 18.60 \tabularnewline
\bottomrule
\end{tabular}} 
\end{table}

\begin{table}
\caption{\label{tab:summary baseline2} Summary statistics of baseline variables (part II)}
\vspace{0.25cm}
\resizebox{0.9\textwidth}{!}{ %
\begin{tabular}{lcc}
\toprule 
\textbf{{n = 1882} }  & \textbf{{Number } }  & \textbf{{Percent } }\tabularnewline
 & \textbf{of Responses}  & \textbf{of Responses (\%)}\tabularnewline
\midrule 
\multicolumn{3}{l}{\textbf{{Health status: Would you say your own health in general
is}}}\tabularnewline
\midrule 
Poor  & 31  & 1.65 \tabularnewline
Fair  & 131  & 6.96 \tabularnewline
Good  & 412  & 21.89 \tabularnewline
Very Good  & 665  & 35.33 \tabularnewline
Excellent  & 318  & 16.90 \tabularnewline
N/A  & 325  & 17.27 \tabularnewline
\midrule 
\multicolumn{3}{l}{\textbf{{Common smoke: Smoke from wildfires is a common occurrence
where I live}}}\tabularnewline
\midrule 
Strongly disagree  & 72  & 3.83 \tabularnewline
Somewhat disagree  & 112  & 5.95 \tabularnewline
Neither agree nor disagree  & 129  & 6.85 \tabularnewline
Somewhat agree  & 575  & 30.55 \tabularnewline
Strongly agree  & 631  & 33.53 \tabularnewline
N/A  & 363  & 19.29 \tabularnewline
\midrule 
\multicolumn{3}{l}{\textbf{{Smoke impact: A few hours of wildfire smoke in the air can
impact my health} }}\tabularnewline
\multicolumn{3}{l}{\textbf{(18 users have two different responses)}}\tabularnewline
\midrule 
Strongly disagree  & 13  & 0.69 \tabularnewline
Somewhat disagree  & 34  & 1.81 \tabularnewline
Neither agree nor disagree  & 131  & 6.96 \tabularnewline
Somewhat agree  & 380  & 20.19 \tabularnewline
Strongly agree  & 958  & 50.90 \tabularnewline
N/A  & 366  & 19.45 \tabularnewline
\midrule 
\multicolumn{3}{l}{\textbf{{Possible to reduce smoke: It is possible for me to reduce
my wildfire smoke}}}\tabularnewline
\midrule 
Strongly disagree  & 45  & 2.39 \tabularnewline
Somewhat disagree  & 129  & 6.85 \tabularnewline
Neither agree nor disagree  & 189  & 10.04 \tabularnewline
Somewhat agree  & 681  & 36.18 \tabularnewline
Strongly agree  & 471  & 25.03 \tabularnewline
N/A  & 367  & 19.50 \tabularnewline
\midrule 
\multicolumn{3}{l}{\textbf{{Alert help: Information alerts are likely to help me reduce
my exposure to wildfire smoke}}}\tabularnewline
\midrule 
Strongly disagree  & 9  & 0.48 \tabularnewline
Somewhat disagree  & 27  & 1.43 \tabularnewline
Neither agree nor disagree  & 99  & 5.266 \tabularnewline
Somewhat agree  & 476  & 25.29 \tabularnewline
Strongly agree  & 897  & 47.66 \tabularnewline
N/A  & 374  & 19.87 \tabularnewline
\bottomrule
\end{tabular}} 
\end{table}

\begin{table}
\caption{\label{tab:summary baseline3} Summary statistics of baseline variables (part III)}
\vspace{0.25cm}
\resizebox{0.9\textwidth}{!}{ %
\begin{tabular}{lcc}
\toprule 
\textbf{{n = 1882} }  & \textbf{{Number} }  & \textbf{{Percent } }\tabularnewline
 & \textbf{of Responses}  & \textbf{of Responses (\%)}\tabularnewline
\midrule 
\multicolumn{3}{l}{\textbf{{Symptoms 1: Do you commonly experience any of these symptoms?
(Select all that apply)}}}\tabularnewline
\midrule 
Coughing, trouble breathing, wheezing, asthma attacks, or similar  & 381  & 20.24 \tabularnewline
High blood pressure, chest pain or tightness, rapid or irregular heartbeat,
or similar  & 156  & 8.29 \tabularnewline
Runny or stuffy nose, irritated sinuses, or similar  & 634  & 33.69 \tabularnewline
Stinging eyes, scratchy throat, or similar  & 429  & 22.79 \tabularnewline
Tiredness, headaches, or similar  & 553  & 29.38 \tabularnewline
None of the above  & 548  & 29.12 \tabularnewline
N/A  & 368  & 19.55 \tabularnewline
\midrule 
\multicolumn{3}{l}{\textbf{{Symptoms 2: Do you commonly experience any of these symptoms?
(Select all that apply)}}}\tabularnewline
\midrule 
Allergies related to the upper respiratory tract eyes and ears  & 566  & 30.07 \tabularnewline
Asthma  & 416  & 22.10 \tabularnewline
Chronic Obstructive Pulmonary Disease  & 43  & 2.28\tabularnewline
Hypertension or high blood pressure  & 241  & 12.81 \tabularnewline
None of the above  & 566  & 30.07 \tabularnewline
Other chronic disease  & 198  & 10.52 \tabularnewline
Other heart disease  & 44  & 2.34 \tabularnewline
Other respiratory disease  & 65  & 3.45 \tabularnewline
Type II diabetes metabolic syndrome or obesity  & 111  & 5.90 \tabularnewline
N/A  & 367  & 19.50 \tabularnewline
\bottomrule
\end{tabular}} 
\end{table}

\subsection{Sensitivity analyses for Smoke Sense data}\label{supp:sensitivity_trt_model}
Note that the dataset presents a substantial proportion (around 50\%) of zero-valued outcomes, which motivates the use of the zero-inflated Poisson regression. When the fitted zero-inflated (ZI) model is substituted for the nuisance outcome function ${\mu}_{m,k}(\overline{L}_m)$, the estimates are $(\widehat{\delta}_{\rm ZI}, \widehat{\mu}_{\rm ZI},\widehat{\sigma}^2_{\rm ZI})=(-0.59,6.5,2.3)$ under the Gaussian distributional-based treatment effect model (\ref{revision-eq:GaussianM}), which is similar to the global estimates $(\widehat{\delta},\widehat{\mu},\widehat{\sigma}^{2})=(-0.3,7.0,3.0)$ in Section \ref{revision-subsec:Global-estimation} of the paper. For the sake of computation time, we choose the regular linear regression for the local estimation and present the associated model diagnosis results in Section \ref{supp:model_diag}.

Besides, instead of using the mean imputation for baseline covariates and last observations carried forward (LOCF) for the time-varying covariates as the main paper, we consider using the multiple imputations (MI) to fill in the missing data. The post-MI estimates are $(\widehat{\delta}_{\rm MI}, \widehat{\mu}_{\rm MI},\widehat{\sigma}^2_{\rm MI})=(-0.34,6.9,2.9)$. As displayed in Figure \ref{fig:glob_gamma}, a similar trend of global treatment effects can be observed.

Furthermore, instead of using Gaussian distributional-based treatment effect model as the paper, we consider a Gamma distributional-based model as follows:
\begin{equation}
    \gamma_{m, k}(\psi^*) = \psi^* (t_m-t_k)^7 \exp\{-(t_m-t_k)\}.
\end{equation}
We obtain the estimated $\widehat{\psi}^*=-4.16\cdot 10^{-4}$. The global treatment effect curve, as presented in Figure \ref{fig:glob_gamma}, the largest magnitude treatment effect is $-0.312$ at lag = 7 months, which is similar to the lag estimate obtained in the Gaussian model. 
\begin{figure}[!ht]
    \centering
    \includegraphics[width=.75\linewidth]{figures/global_est_trt_all.pdf}
    \caption{\label{fig:glob_gamma}Sensitivity analyses of the estimated global treatment effect curve with different imputation method and functional form of treatment effects.}
\end{figure}

\subsection{\label{sec:results_more} Results of $\psi^{*}(\bs^{*})$ estimation at multiple locations
and their confidence intervals}

As a supplement for Section \ref{revision-subsec:Local-constant-estimation},
we present the complete results of point estimates of $\psi^{*}(\bs^{*})=\{\delta(\bs^{*}),\mu(\bs^{*}),\sigma(\bs^{*})^2\}$
and the corresponding confidence intervals for the selected 28 locations
in Table \ref{tab:point_ests}. Moreover, we provide the well-tuned bandwidth $\tau_{\text{cv}}$ for the selected locations, cities, and national parks in Figure \ref{fig:tau_real_data}. The selected $\tau_{\text{cv}}$ have similar ranges as those in (S3) and (S4) of the simulation studies. Besides, we observe that the selected bandwidth $\tau_{\text{cv}}$ in the southwest locations is relatively larger than those associated with the users in the northwest. In light of this finding, one can conclude that the spatial treatment effect function is relatively flat in the southwest but is steeper in the northwest due to the smaller bandwidth window $\tau_{\text{cv}}$.

\begin{table}
\caption{\label{tab:point_ests} Parameter estimates with bias correction and
confidence intervals for the 28 locations with their latitude and
longitude. ''State'' is the state to which each location belongs
or is closest. And the number of bootstrap samples
used to construct CIs is 200. }
\vspace{0.25cm}
\resizebox{\textwidth}{!}{{%
{
\begin{tabular}{cccccc}
\toprule 
Latitude  & Longitude  & State  & $\widehat{\delta}$ (CI)  & $\widehat{\sigma}^{2}$ (CI)  & $\widehat{\mu}$ (CI) \tabularnewline
  \hline
47.0 & -123.0 & Washington & -2.2 ( -3.9, -0.6) &  3.3 ( 2.7,  3.9) & 6.6 ( 5.9,  7.3) \\ 
 47.0 & -121.9 & Washington &  1.4 ( -3.4,  6.2) &  1.5 (-4.4,  7.4) & 7.8 ( 5.8,  9.8) \\ 
47.0 & -120.7 & Washington & -2.2 ( -3.5, -0.8) &  3.2 ( 2.9,  3.6) & 6.6 ( 6.2,  7.0) \\ 
47.0 & -119.6 & Washington & -0.6 ( -4.5,  3.2) &  2.9 ( 1.3,  4.6) & 7.1 ( 5.2,  9.0) \\ 
47.0 & -118.4 & Washington & -2.3 ( -3.6, -1.0) &  3.2 ( 2.5,  3.9) & 6.7 ( 6.2,  7.3) \\ 
47.0 & -117.3 & Washington & -0.5 ( -3.4,  2.5) &  3.1 ( 1.1,  5.1) & 6.9 ( 5.3,  8.5) \\ 
47.0 & -116.1 & Idaho & -1.5 ( -4.0,  1.1) &  3.1 ( 1.5,  4.7) & 6.9 ( 5.4,  8.3) \\ 
44.7 & -123.0 & Oregon & -5.0 (-16.2,  6.2) &  2.7 ( 0.7,  4.7) & 8.5 ( 4.9, 12.0) \\ 
44.7 & -121.9 & Oregon & -3.6 ( -5.9, -1.4) &  3.4 ( 2.5,  4.2) & 6.4 ( 5.1,  7.6) \\ 
44.7 & -120.7 & Oregon & -2.9 ( -6.0,  0.2) &  3.0 ( 2.2,  3.8) & 5.9 ( 4.3,  7.6) \\ 
42.3 & -120.7 & Oregon &  0.0 ( -2.0,  2.0) &  3.0 ( 0.1,  5.9) & 7.0 ( 3.3, 10.7) \\ 
42.3 & -119.6 & Oregon &  0.2 ( -2.5,  2.9) & -0.1 (-2.9,  2.7) & 8.9 ( 5.4, 12.5) \\ 
42.3 & -118.4 & Oregon & -0.8 ( -3.6,  2.1) &  3.6 ( 0.9,  6.3) & 6.4 ( 2.4, 10.4) \\ 
40.0 & -123.0 & California &  1.2 (  0.5,  2.0) &  3.0 ( 2.2,  3.8) & 7.0 ( 6.3,  7.6) \\ 
40.0 & -121.9 & California & -0.8 ( -1.3, -0.3) &  3.0 ( 2.6,  3.3) & 7.1 ( 6.9,  7.2) \\ 
40.0 & -120.7 & California &  0.2 ( -6.7,  7.1) &  3.0 (-6.3, 12.3) & 7.1 ( 3.6, 10.6) \\ 
40.0 & -119.6 & Nevada & -1.7 ( -6.5,  3.2) &  2.4 (-0.7,  5.5) & 6.3 ( 4.2,  8.4) \\ 
37.7 & -123.0 & California &  0.6 (  0.2,  1.0) &  3.0 ( 1.8,  4.2) & 7.0 ( 6.6,  7.4) \\ 
37.7 & -121.9 & California & -0.7 ( -1.0, -0.4) &  2.9 ( 2.8,  3.1) & 7.1 ( 7.0,  7.2) \\ 
37.7 & -120.7 & California & -1.9 ( -2.8, -1.1) &  3.2 ( 2.8,  3.6) & 6.7 ( 5.4,  8.0) \\ 
37.7 & -119.6 & California & -4.0 (-13.5,  5.6) &  2.9 (-0.3,  6.1) & 5.2 ( 0.3, 10.1) \\ 
35.3 & -119.6 & California & -3.0 ( -9.2,  3.2) &  2.9 (-0.5,  6.4) & 6.0 ( 2.4,  9.6) \\ 
35.3 & -118.4 & California &  7.4 ( -4.7, 19.5) &  3.2 (-0.3,  6.7) & 5.2 (-2.6, 13.0) \\ 
35.3 & -117.3 & California & -1.5 ( -2.2, -0.8) &  3.1 ( 2.6,  3.6) & 6.9 ( 5.6,  8.1) \\ 
35.3 & -116.1 & California & -1.6 ( -2.4, -0.9) &  3.1 ( 2.4,  3.8) & 6.8 ( 5.4,  8.2) \\ 
33.0 & -118.4 & California & -1.4 ( -2.2, -0.6) &  3.1 ( 2.6,  3.5) & 6.9 ( 6.5,  7.3) \\ 
33.0 & -117.3 & California & -1.6 ( -2.4, -0.8) &  3.1 ( 2.5,  3.7) & 6.8 ( 6.2,  7.5) \\ 
33.0 & -116.1 & California & -2.0 ( -3.1, -0.9) &  2.4 ( 1.8,  3.1) & 6.6 ( 5.8,  7.4) \\ 
   \hline
\end{tabular}} } }
\end{table}

\begin{figure}[!ht]
    \centering
    \includegraphics[width=.8\linewidth]{figures/selected_tau_real_data.pdf}
\caption{\label{fig:tau_real_data}Optimal $\tau_{\text{cv}}$ for the selected 28 locations in the Western U.S., cities, and national parks. $\tau_{\text{cv}}$ is the tuning parameter chosen
by the $4$-fold cross-validation. }
\end{figure}

\subsection{Model diagnoses for assessment\label{supp:model_diag}}
In this section, we first assess the model fits for the outcome mean $\mu_{m,k}(\overline{L}_m)$ and the propensity score $e(\overline{L}_m)$. Figure \ref{fig:residual} showcases the residual error plots for these two models. Notably, the residuals are evenly distributed around zero, indicating satisfactory model fits.
\begin{figure}[!ht]
    \centering
    \includegraphics[width=.8\linewidth]{figures/residual_ps_om.pdf}
    \caption{Residual error plots for the nuisance functions: (left) the propensity score $e(\overline{L}_m)$ and (right) the outcome mean $\mu_{m,k}(\overline{L}_m)$.}
    \label{fig:residual}
\end{figure}

Furthermore, we propose a model diagnostic statistic to measure how well the proposed treatment effect model fits the data. In essence, if the models $\gamma_{m,k}\{\psi^*(\boldsymbol{s})\}$, $\widehat{\mu}_{m,k}(\overline{L}_m)$ and $\widehat{e}(\overline{L}_m)$ are well-fitted, $Y_k 
    - \sum_{l=m}^k 
    \gamma_{l,k}\{\widehat{\psi}(\boldsymbol{s}^*)\}-\widehat{\mu}_{m,k}(\overline{L}_m)$ should be approximately uncorrelated with $A_m-\widehat{e}(\overline{L}_m)$ given $\overline{L}_m$. Therefore, a model diagnostic statistic $r\{V;\psi(\boldsymbol{s}^*)\}$ can be formulated:
\begin{align*}
        &r\{V;\psi(\boldsymbol{s}^*)\}\\
        &=\mathbb{P}_n
    \omega_{\tau}(\|\boldsymbol{s}^*-\boldsymbol{s}\|)
    \\
    &
    \times
    \sum_{m=1}^K 
    \sum_{k=m}^K
    h_{m,k}(\overline{L}_m)
    \left[ 
    H_k^{(\overline{A}_{m-1})}\{\widehat{\psi}(\boldsymbol{s}^*)\}
    - \widehat{\mu}_{m,k}(\overline{L}_m)
    \right]
    \{
    A_m - \widehat{e}(\overline{L}_m)
    \}\\
    &=\mathbb{P}_n 
    \omega_{\tau}(\|\boldsymbol{s}^*-\boldsymbol{s}\|)
    \\
    &\times
    \sum_{m=1}^K 
    \sum_{k=m}^K
    h_{m,k}(\overline{L}_m)
    \left[ 
    Y_k 
    - \sum_{l=m}^k 
    \gamma_{l,k}\{\widehat{\psi}(\boldsymbol{s}^*)\}
    - \widehat{\mu}_{m,k}(\overline{L}_m)
    \right]
    \{
    A_m - \widehat{e}(\overline{L}_m)
    \},
    \end{align*}
    for a set of functions $\{h_{m,k}(\overline{L}_m), m=1,\cdots,K, k = m,\cdots,K\}$. The functional form of $h_{m,k}(\overline{L}_m)$ is motivated by the domain-specific knowledge about the strength of the measured covariates.

    In our wild bootstrap method for variance estimation, we generate the exchangeable random weights $\xi_i, i=1,\cdots,n$. Thus, we can re-compute the estimate $r\{V;\psi(\boldsymbol{s}^*)\}$ using weighted analysis with $\xi_i$ for each bootstrap replication, which does not pose additional computational burdens. Assisted by the bootstrapped estimates $r^{(b)}\{V;\psi(\boldsymbol{s}^*)\}, b=1,\cdots, B$, we can construct the Wald-type confidence interval for each location $\boldsymbol{s}^*$.

    After querying domain experts for the important confounders, "days feeling smoke inside" and "days feeling smoke outside" are two critical time-varying variables in their judgment. Hence, the functional form of $h_{m,k}(\overline{L}_m)$ is determined according to these two variables, and the associated confidence intervals are constructed adjusting for multiple comparisons. The forest plots in Figure \ref{fig:model_diag_locs} and \ref{fig:model_diag_city} suggest that there is no significant evidence to reject the conditional independence between $H_k^{\overline{A}_{m-1}}\{\widehat{\psi}(\boldsymbol{s}^*)\}$ and $A_m-\widehat{e}(\overline{L}_m)$ for most selected locations and cities, and thus conclude satisfactory model fits. 
    \begin{figure}[!ht]
        \centering\includegraphics[width=.9\linewidth]{figures/g_loss_ci_locs_cov1.pdf}
        \includegraphics[width=.9\linewidth]{figures/g_loss_ci_locs_cov2.pdf}\caption{\label{fig:model_diag_locs}Estimated model diagnostic statistic $r\{V;\psi(\boldsymbol{s}^*)\}$ and its 95\% confidence intervals (adjusting for multiple comparisons) for the selected locations.}
    \end{figure}  
    \begin{figure}[!ht]
        \centering \includegraphics[width=.8\linewidth]{figures/g_loss_ci_city_cov1.pdf}
        \includegraphics[width=.8\linewidth]{figures/g_loss_ci_city_cov2.pdf}\caption{\label{fig:model_diag_city}Estimated model diagnostic statistic $r\{V;\psi(\boldsymbol{s}^*)\}$ and its 95\% confidence intervals (adjusting for multiple comparisons) for the selected cities and national parks.}
    \end{figure}



\newpage
\bibliographystyle{chicago}
\bibliography{refs}